\documentclass[lettersize,journal]{IEEEtran}
\usepackage{diagbox}
\usepackage{cite}
\usepackage{graphicx} % Required for inserting images
\usepackage[inline]{enumitem}
\usepackage[most]{tcolorbox}
\usepackage{color}
\definecolor{pblue}{rgb}{0.13,0.13,1}
\definecolor{pgreen}{rgb}{0,0.5,0}
\definecolor{pred}{rgb}{0.9,0,0}
\definecolor{pgrey}{rgb}{0.46,0.45,0.48}
\usepackage{listings}
\usepackage{hyperref}
\usepackage{amsmath,amsfonts}
\usepackage{multirow}
\usepackage{subcaption}
\usepackage{threeparttable}
\usepackage{balance}
\usepackage{enumerate}
\usepackage{booktabs}

\newcommand{\project}[1]{{\textit{#1}}}
\newcommand{\code}[1]{{\texttt{#1}}}

\usepackage{boxedminipage}
\newenvironment{result}%
{\smallskip
   \noindent
   \let\emph=\textbf
   \begin{boxedminipage}{\columnwidth}\em\textbf{\textsc{Conclusion:}}}%
      {\end{boxedminipage}%
   \smallskip
}

\lstdefinestyle{Java}{
    language={Java}, basicstyle=\footnotesize,
    moredelim=[is][\color{red}]{@@}{@@}
}

\title{Demystifying and Assessing Code Understandability in Java Decompilation}

\author{\IEEEauthorblockN{
        Ruixin Qin\IEEEauthorrefmark{1}\thanks{*Both authors contributed equally to this paper.}
        Yifan Xiong\IEEEauthorrefmark{1},
	Yifei Lu\IEEEauthorrefmark{2}\thanks{\IEEEauthorrefmark{2}Corresponding authors.}, 
	Minxue Pan\IEEEauthorrefmark{2}}
    
	\IEEEauthorblockA{State Key Laboratory for Novel Software Technology, Software Institute, Nanjing University, China}
 
 \IEEEauthorblockA{Email: ruixinqin@smail.nju.edu.cn, 211250043@smail.nju.edu.cn, lyf@nju.edu.cn, mxp@nju.edu.cn}
} 

\begin{document}

\maketitle

\begin{abstract}
Decompilation, the process of converting machine-level code into readable source code, plays a critical role in reverse engineering. Given that the main purpose of decompilation is to facilitate code comprehension in scenarios where the source code is unavailable, the understandability of decompiled code is of great importance. Unfortunately, previous researches have predominantly concentrated on the correctness of decompilation, leaving the understandability of the decompiled code largely unexplored. Do decompiler stakeholders place importance on the understandability of decompiled code? Are there any methodologies that can be used to assess this understandability? These questions, however, remain unanswered so far.

Therefore, in this paper, we propose the first empirical study on the understandability of Java decompiled code. This study involves a well-designed user survey to reveal the Severity and Universality of understandability issues in Java decompilation, as well as a series of experiments for the understandability comparison between a total of 429 sets of source code files from 14 Java projects and corresponding decompiled files provided by 3 famous Java decompilers. Through an in-depth analysis of the survey results and the experiment results, we obtained the following findings:
\begin{enumerate*}
    \item[(1)] Understandability of Java decompilation is considered as important as its correctness, and decompilation understandability issues are even more commonly encountered than decompilation failures.
    \item[(2)] A notable percentage of code snippets decompiled by Java decompilers exhibit significantly lower or higher levels of understandability in comparison to their original source code.
    \item[(3)] Unfortunately, Cognitive Complexity demonstrates relatively acceptable precision while low recall in recognizing these code snippets exhibiting diverse understandability during decompilation. 
    \item[(4)] Even worse, perplexity demonstrates lower levels of precision and recall in recognizing such code snippets.
    % \item[(1)] Cognitive Complexity is effective in evaluating the complexity of code structure in the context of decompilation.
    % \item[(2)] Perplexity is able to capture uncommon patterns of decompiled code from the perspective of expressions, formatting and constants to a limited extent.
    % \item[(3)] Understandability of decompiled code is no better than the original counterpart in general, while there exists a few exceptions. 
    % \item[(4)] Common patterns that influence code understandability are observed across decompilers.
\end{enumerate*}
Inspired by the four findings, we further proposed six code patterns that are popular among real-world Java projects and are highly possible to confuse existing metrics in recognizing the understandability diversity between decompiled and source code. Based on the six code patterns, the first metric for the assessment of decompiled code understandability was proposed. This metric was extended from Cognitive Complexity, with six more rules harvested from an exhaustive manual analysis into 1287 pairs of source code snippets and corresponding decompiled code. This metric was also validated using the original and updated dataset, yielding an impressive macro F1-score of 0.88 on the original dataset, and 0.86 on the test set.

%In this paper, we strive to assess the understandability of decompiled code by employing Cognitive Complexity and perplexity of n-gram language models. Our empirical findings reveal that neither of these metrics performs well. 
% To propose a better metric, we conduct an exhaustive manual analysis, in which we meticulously compare 1287 pairs of decompiled source code and its original counterpart. Through this labor-intensive analysis, we identify six recurring code patterns that impact the understandability of decompiled code. These patterns can be categorized into four aspects: Code Structure, Expression, Formatting, and Constants. In response to these observations, we introduce a novel metric based on Cognitive Complexity, specifically tailored for the assessment of decompiled code understandability. 

\end{abstract}


% 20231218 from the perspective of users ?
\section{Introduction} \label{sec:intro}
Java decompilation, the process of converting low-level Java bytecode back into human-readable source code, has been widely used in scenarios involving reverse engineering and legacy code maintenance. A number of studies~\cite{harrand2020java,mauthe2021large, liu2023empirical,lu2024} have been conducted to illustrate the latest progresses of the decompilation for various languages with the goal of demystifying the inherent challenges within this domain. However, none of these research works have ever studied the understandability of decompiled code, or decompilation understandability for short. 
%, and particularly, the understandability disparities between decompiled code and their corresponding source code. 

Java source code, being the human-readable representation of a program, is known for its expressive and high-level constructs that closely reflect the programmer's logic and intentions. However, decompilers may occasionally generate difficult-to-read code snippets that significantly diverge from the elegance and understandability characteristic inherent in the source code. % to fix
The ramifications of such divergence extend beyond mere inconvenience, impacting the overarching objective of program comprehension intrinsic to Java decompilation. Indeed, the prevalence of these understandability-related issues undermines the user experience and, by extension, impedes the broader dissemination of Java decompilers. An observable trend indicates a rise in the frequency of reported comprehension issues, indicative of heightened scrutiny from both users and developers on the understandability of Java decompilation.
Illustratively, the widely utilized Java decompiler project, \project{Jadx}, renowned for its substantial user base and active community engagement (over 39,000 stars and 1,300 issues on GitHub), has garnered attention for its proportionally significant volume of understandability-related issues. Specifically, out of 159 issues concerning decompiled code generation reported within the past three years, over 21\% are related to decompilation understandability, marking a discernible escalation in comparison to historical data.
Moreover, over 53\% of these issues have already been fixed, underscoring the common commitment from both users and developers aimed at improving the understandability of Java decompilation.
%These issues hinder the widespread adoption and the further advancement of decompilation technology, therefore both users and developers are motivated to report and address these issues to ensure the continued development of Java decompilation techniques.
%The increasing number of reported comprehension issues imply that both decompiler users and developers are paying increasing attentions on the understandability of decompilation results.

However, prior to enhancing the decompilation understandability, it is imperative to conduct an objective and accurate assessment of decompilation understandability. 
Unfortunately, it is a non-trivial task, as understandability is a relative concept. The majority of insights into understandability-related issues are observed from individual usage experiences, lacking stringent standards. Moreover, no previous researches have ever systematically delved into the understandability of Java decompilation. 
Though several rule-based metrics such as Cognitive Complexity~\cite{campbell2018cognitive} and machine-learning-based metrics like perplexity~\cite{manning1999foundations} are employed in source code comprehension tasks, their application in the context of Java decompilation remains unexamined. The viability of employing these metrics to assess decompilation understandability, as well as the selection of code snippets as the understandability standard for assessment, warrants further investigation.

Moreover, assessing decompilation understandability should also be a long-term task as well. Being compiled into bytecode, Java source code may have quite a number of details dropped. This fact necessitates developers to judiciously employ heuristic strategies in selecting the most easily understandable code structures to generate. However, due to the limited generalizability and susceptibility to inherent bias, these heuristic strategies may typically apt only for specific code patterns, thereby diminishing the understandability of other scenarios. 
% [core] Reduce the number embedded blocks #1455 https://github.com/skylot/jadx/issues/1455
% [core] Weird if conditions #2 #1689 https://github.com/skylot/jadx/issues/1689
Still in \project{Jadx}, developers first tried to fix the nested code structure case~\footnote{\url{https://github.com/skylot/jadx/issues/1455}} as shown in \autoref{lst:nested-a}. However, as indicated by a later issue~\footnote{\url{https://github.com/skylot/jadx/issues/1689}}, the proposed fix only work on a small portion of cases while \project{Jadx} still generate code as shown in \autoref{lst:nested-b}.
Such enduring challenge highlight the prospect of continual emergence of understandability-related issues, which have plagued both decompiler developers and users. Therefore, there exists an urgent demand of efficient assessment approach for the understandability of Java decompilation, which help developers continuously improve their decompilers.

\begin{figure}[htbp]
\centering
    \begin{subfigure}{.22\textwidth}
        \begin{lstlisting}[style=Java, basicstyle=\scriptsize]
if (condition1) {
    if (condition2) {
        if (condition3) {
            ...
        } else {...
        }
    } else {...
    }
} else {...
}
        \end{lstlisting}
    \vspace{0.1cm}
        \caption{Code snippet 1}\label{lst:nested-a}
    \end{subfigure}
    \hspace{0.1cm}
    \begin{subfigure}{.22\textwidth}
        \begin{lstlisting}[style=Java, basicstyle=\scriptsize]
if (condition1) {
    if (condition2) {
        if (condition3) {
            return null;
        }
        return ...;
    }
    return ...;
}
return ...;
        \end{lstlisting}
    \vspace{0.1cm}
        \caption{Code snippet 2}\label{lst:nested-b}
    \end{subfigure}
\caption{Excerpts of nested code structures generated by \project{Jadx}}
\label{fig:nested-all}
\end{figure}

Therefore, in this paper, we conduct an empirical study on code understandability of Java decompilation, which is organized by following the four research questions:

% Inspired by this questionnaire, we plan to access the from a standard and quantified manner. 
% Afterwards, we conduct an empirical study on code understandability of Java decompilation, which is motivated by following four research questions:

\textbf{RQ1: How do stakeholders in Java decompilation perceive and prioritize the understandability of decompiled code?}
While the proliferation of understandability-related issues offers an indirect evidence of the growing emphasis placed on code understandability within Java decompilation, an accurate assessment of the extent to which pertinent stakeholders, such as Java program developers, Java decompiler users, and researchers, actively prioritize decompilation understandability necessitates a direct inquiry via survey.

\textbf{RQ2: how can code understandability vary between decompiled code and source code?} 
The major concern of the code understandability in decompilation lies in whether Java decompilers are able to retain the understandability of the code. This prompts an in-depth investigation into whether decompiled code may exhibit different understandability compared to its source code counterpart, and the underlying factors contributing to such disparities.

\textbf{RQ3: how can rule-based understandability metrics assess the understandability of both decompiled code and their corresponding source code?} 
Conventional metrics, such as the widely recognized Cognitive Complexity, which are rule-based in nature, serve as established measures for assessing the understandability of provided segments of source code\cite{munoz2020empirical}. An interesting topic concerns the applicability of these metrics in the context of decompilation. 

\textbf{RQ4: how can language-model-based metrics assess the understandability of both decompiled code and their corresponding source code?} 
Language models leveraging real-world projects for intelligence, notably perplexity, typically serve as a guiding framework to enhance the code understandability in program completion and generation tasks~\cite{svyatkovskiy2020intellicode}. Consequently, we also examine the applicability of perplexity as an understandability metric of decompiled code.

To address the research questions posed above, a comprehensive research methodology including a user survey and three experiments, has been proposed in this paper, which gives both a macro and a micro picture of the understandability of Java decompilation, especially, in a quantitative manner.
Our user survey targeting the stakeholders' perspectives on Java decompilation understandability has revealed that Java developers and researchers view the understandability of Java decompilation as equally important as its correctness, with understandability issues occurring more frequently than decompilation failures. 
We also propose the first series of experiments for Java decompilation understandability, in which the understandability of Java decompiled code is assessed in comparison to its corresponding source code, according to which the results are classified into three types, namely \textbf{\textit{Less}}, \textbf{\textit{Equi}}, and \textbf{\textit{More}}. During these experiments, six code patterns have been revealed, which contribute to major disparities in code understandability between decompiled code and source code. 
Building upon these code patterns, we propose and subsequently validate a comprehensive metric based on Cognitive Complexity, namely \emph{Cognitive Complexity\textsuperscript{D}}, to assess understandability of decompiled code. The experimental results prove the effectiveness of this metric in accurately identifying the understandability of decompiled code. Particularly, This metric exhibits notable recall and precision in identifying disparities in comprehensibility between decompiled code and its source code counterpart.
%Although neither of the two metrics provides a comprehensive evaluation of understandability of decompiled code, Cognitive Complexity is effective in capturing the complexity of code structure, and perplexity offers insights into a few uncommon patterns. Building upon these valuable insights, we propose and subsequently validate a comprehensive metric based on Cognitive Complexity to assess understandability of decompiled code.

% This metric serves as a valuable contribution to the field, offering a multifaceted approach to evaluating code understandability in the context of Java decompilation.

\subsection{Contribution}
We summarize our contributions as follows:
\begin{itemize}
    \item a well-designed user survey on decompilation that investigates stakeholders' perspectives on the understandability of Java decompilation.
    \item a ground-truth benchmark for code understandability assessment in the context of decompilation, comprising a total of 1287 files sourced from 14 Java projects and their counterparts resulting from the compilation-decompilation process.
    \item the first in-depth empirical study on code understandability of Java decompilation, in which six code patterns that greatly confuse the two understandability metrics were summarized from this study.
    \item an enhanced understandability metric inspired by the six code patterns for the context of decompilation and a tool to assess understandability for given pieces of code with this enhanced metric.
    \item a series of experiments on this enhanced metric  to evaluate its effectiveness in revealing decompiled code snippets manifesting clearly distinguishable levels of understandability.
\end{itemize}

\subsection{Paper Organization.}

The rest of the paper is organized as follows: \autoref{sec:survey} describes our questionnaire results and demonstrates the magnitude of our study. 
\autoref{sec:background} introduces the understandability metrics used in our study. \autoref{sec:methodology} introduces our experimental methodology. \autoref{sec:results} describes our experimental results. \autoref{sec:casestudy} describes our case studies.
\autoref{sec:new_metric} designs and validates our new metric. 
\autoref{sec:threat} discusses the threats. \autoref{sec:related_work} introduces the related work. \autoref{sec:conclusion} concludes the paper. \autoref{sec:dataavailable} provides the link to our experimental data and tools. 

\section{Magnitude}
\label{sec:survey}

% First, 
To provide a macro picture of the understandability of Java decompilation, a well-designed user survey has been proposed, which targets stakeholders engaged in Java development regarding the understandability of Java decompilation. In this survey, a questionnaire, standing as one of the most effective and frequently-used methods~\cite{DBLP:journals/ase/HuangHCHZ23, enders2023dewolf, dos2018impacts} for collecting direct feedback from users, has been introduced. This questionnaire enables us to discern prevalent commonalities and disparate perspectives among diverse stakeholders from a macroscopic viewpoint, thereby enriching comprehension of the nature characteristics of Java decompilation. In a nutshell, this user survey aims to systematically and directly address our RQ1.
%A questionnaire is one of the most efficient way to collect the most direct reflection from Java stakeholders on the Java decompilation understandability, and therefore is widely used\cite{DBLP:journals/ase/HuangHCHZ23, enders2023dewolf, dos2018impacts}. This allows researchers to identify common themes and divergent viewpoints among stakeholders, enhancing understanding of the nature of understandability of Java decompilation.
%In a net shell, this questionnaire is meticulously designed to probe stakeholders' perceptions and priorities regarding Java decompilation, thereby directly addressing our RQ1. 
\subsection{Setup For User Survey}
The setup of our user survey for RQ1 is as follows.

\subsubsection{Survey Participants}

% participants with different backgrounds
Our target participants include developers of Java decompilers, developers of popular Java open-source projects and academic researchers.
Developers of decompilers possess specialized knowledge and experience in decompilation techniques, providing valuable insights into tool development and usage. Engaging with developers of open-source projects allows us to capture the perspectives of potential end users of Java decompilers in real-world software development scenarios. % todo should we describe the reasons of choosing academic researchers?
In detail, 234 developers from 8 well-known Java decompiler projects on GitHub, 451 developers from 10 popular Java projects, and 120 researchers majored in computer science from 3 world-famous universities have been selected as our target participants. Eventually, 805 developers and researchers were invited to participate in this user survey in total.
% We sent the link of the survey to the participants. 
% We sent the questionnaire to 234 developers of Java decompilers and received 15 responces, 650 developers of popular Java open-source projects and received 10 responces. We also asked ten students majoring in Computer Scienct to independently complete the survey. We asked ten graduate students to independently complete the survey
% , offering practical insights into usability and effectiveness.

\subsubsection{Survey Platform}

E-mails have been employed to gather questionnaire responses from participants across different regions and backgrounds, during which anonymity is ensured. The e-mail addresses of our target participants were collected by examining the commit logs of well-known Java projects and decompiler projects, as well as the contact information available on the websites of universities.
% We employed commit logs of open-source projects in GitHub to find developers' emails and send invitations to them. We also invited academic researchers to complete the survey.

\begin{table}[t]
    \centering
    \caption{Survey Questions}
    \setlength{\tabcolsep}{0.1em}
    \begin{tabular}{|c|>{\centering\arraybackslash}m{3.4cm} |c|>{\centering\arraybackslash}m{3.0cm}|}
        \hline
        \textbf{No.}        & \textbf{Question}     &  \textbf{Type}        & \textbf{Options} \\
        \hline
        \multirow{2}{*}{Q1} & Java decompilation    &                       &  \multirow{2}{*}{Yes, No} \\
                            & experience            &                       & \\
        \cline{1-2} \cline{4-4}
        \multirow{2}{*}{Q2} & \multirow{2}{*}{Education level} &            & PhD, Master, Bachelor, \\
                            &                                  &            & High School\\
        \cline{1-2} \cline{4-4}
                        Q3  &  Java programming experience     &            & 1 year to 10 years\\
        % \multirow{2}{*}{Q3} & \multirow{2}{*}{Java programming experience} & & 1 to 10 \\
        %                     &                                  &            & ... \\
        \cline{1-2} \cline{4-4}
        \multirow{3}{*}{Q4} & \multirow{3}{*}{Java decompilers ever used} &         & CFR, FernFlower, JADX,\\
                            &                                             & Self-   & JD-GUI, Krakatau, \\
                            &                                             &evaluation& Procyon, Other\\
        \cline{1-2} \cline{4-4}
        \multirow{5}{*}{Q5} & \multirow{5}{*}{Java decompilers usage}     &         & Bug Detection, \\
                            &                                             &         & Code Reuse, \\
                            &                                             &         & Malware Detection, \\
                            &                                             &         & Privacy Leakage\\
                            &                                             &         & Detection, Other \\
        \cline{1-2} \cline{4-4}
        \multirow{2}{*}{Q6} & \multirow{2}{*}{Decompiler usage frequency} &         & Daily, Weekly, \\
                            &                                             &         & Monthly, Less\\
        \hline % \multirow{6}{*}{Decompilation }
        \multirow{2}{*}{Q7} &  \multirow{2}{*}{Importance of decompilation}&        & 1 (Not Important) to \\
                            &                                              &        & 10 (Very Important)\\
        \cline{1-2} \cline{4-4}
        \multirow{2}{*}{Q8} & Importance of decompilation           & Decompilation & 1 (Not Important) to \\
                            & correctness                           & evaluation    & 10 (Very Important)\\
        \cline{1-2} \cline{4-4}
        \multirow{2}{*}{Q9} &  Importance of decompilation                  &       & 1 (Not Important) to \\
                            & understandability                             &       & 10 (Very Important)\\
        \hline
        \multirow{2}{*}{Q10}&  Frequency of decompilation           &               & 1 (Very rare) to \\
                            & failures                              & Decompilation & 10 (Very frequent)\\
        \cline{1-2} \cline{4-4}                                     
        \multirow{2}{*}{Q11}&  Frequency of decompilation           & issues        & 1 (Very rare) to \\
                            & understandability issues              &               &  10 (Very frequent)\\
        % \multirow{2}{*}{Q10}&  Frequency of decompilation           &               & 1 (Very rare) to \\
        %                     & failures                              & Frequency of  & 10 (Very frequent)\\
        % \cline{1-2} \cline{4-4}
        % \multirow{2}{*}{Q11}&  Frequency of decompilation           & decompilation & 1 (Very rare) to \\
        %                     & understandability issues              & issues        & 10 (Very frequent)\\
        \hline

        % \hline
    \end{tabular}
    \label{tab:survey}
\end{table}

% type: Decompilation experience   Decompilation difficulty/problems/issues ?  Encountered Issues

% Q1, Q4, Q7 not put
\subsubsection{Survey Questions}
A total of 11 questions have been designed in the questionnaire. The first question, Q1, inquires whether participants have ever used decompilers, which is critical given its potential influence on the reliability of their following responses concerning Java decompilation.
The remaining questions can be classified into three categories: (1) the participants’ self-evaluation, including their education level (Q2), their experience in Java programming (Q3); experience in Java decompilers usage, including the Java decompilers ever used (Q4), the purpose they used the decompilers for (Q5) and the frequency of using decompilers (Q6); (2) the importance of Java decompilation (Q7), decompilation correctness (Q8), and decompilation understandability in the participants' view (Q9); (3) the frequency of decompilation failures (Q10) and understandability issues (Q11) encountered by participants.

Here we collect the detailed information of participants in order to investigate whether the experience and the characteristics of participants may influence their opinion on Java decompilation.
We focus not only on the decompilation understandability but also on the decompilation correctness, as correctness is another important concern in decompilation. It can help us have a better comprehension of the importance of decompilation understandability and frequency of understandability issues in comparison. 
Particularly, question Q1 is designed to help us exclude invalid responses from users who have never used Java decompilers. Questions Q7-Q11 are 10-point Likert scale questions with 10 selections ranging from 1 (not important/very rare) to 10 (very important/very frequent).

\subsection{Survey Results}

\begin{figure}[htbp]
    \centering
    \begin{minipage}[]{\linewidth}
        \includegraphics[width= 0.95\linewidth]{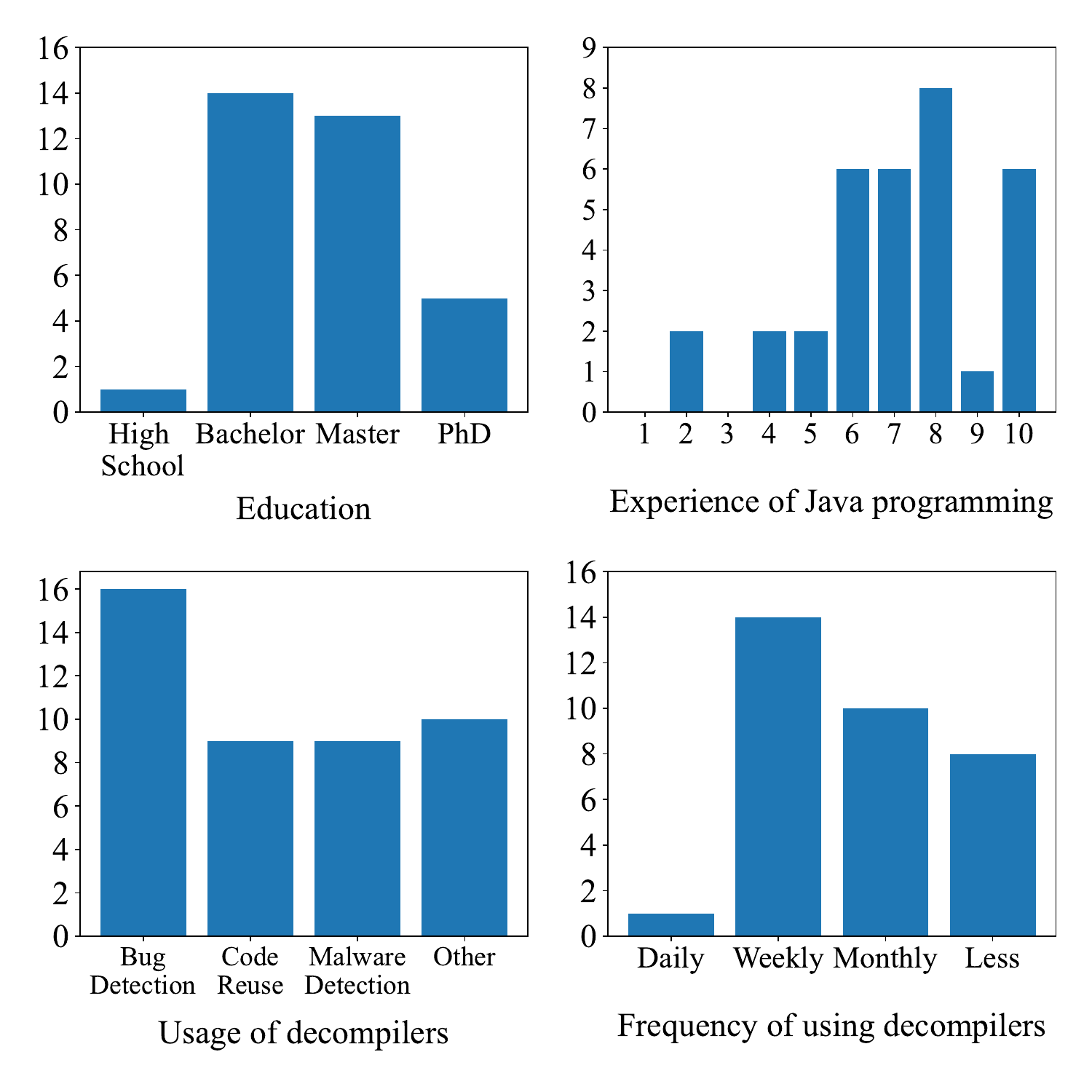}
        \subcaption{Respondents' self-evaluation and Java decompiler experience.}
        \label{fig:survey1}
        
        \includegraphics[width= 0.95\linewidth]{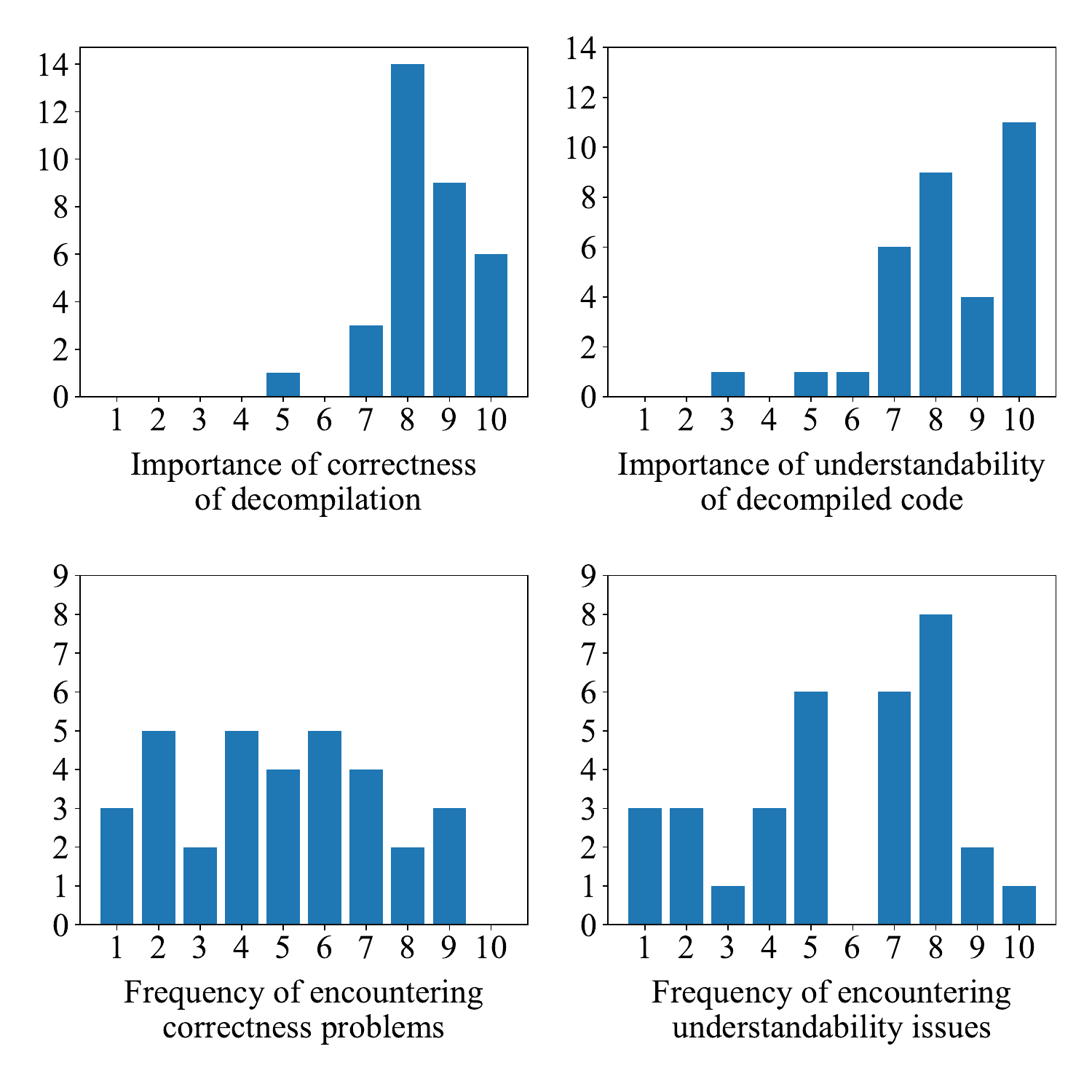}
        \subcaption{Respondents' evaluation on the correctness and understandability of Java decompilation.}
        \label{fig:survey2}
    \end{minipage}
    \caption{Survey results.}
\end{figure}

Eventually, we received 33 valid survey responses, out of which 15 were from developers of Java decompilers, 9 were from developers of other Java projects, and 9 were from university researchers. % Figures~\autoref{fig:survey1} 
\autoref{fig:survey1} presents the respondents' self-evaluation and their usage experience with Java decompilers. Among all 33 respondents, 97\% (32 out of 33) hold at least a bachelor degree, with over half holding a master's degree or a PhD degree. In addition, 82\% respondents (27 of 33) have more than five years of experience in Java programming. These facts indicate that the survey sample consists of respondents with advanced degrees and rich experience in Java programming. These respondents are likely to have a deeper understanding of Java programming and be encountered with a wider diversity of coding styles and practices. Hence, these respondents are better equipped to evaluate the correctness and the understandability of Java code, especially the decompiled code, making their responses more reliable.

% have used any one of the three decompilers.
% 29/33 = 0.8787878787878788
% FernFlower 18+3(idea jetbrain)  JD-GUI 16  Jadx 12  CFR 10
\project{FernFlower}, \project{JD-GUI}, \project{Jadx} and \project{CFR},  are four of the most frequently used decompilers, and over 88\% of respondents have used any one of the four decompilers. % Since \project{JD-GUI} has not been actively maintained since its last update in 2019, we exclude it.
% \project{FernFlower} is the most frequently used decompilers, and over 64\% of respondents have used it.
The purposes for using Java decompilers are quite diverse: the three most common motivations are bug detection (16 respondents), code reuse (9 respondents), and malware detection (9 respondents). Additionally, several other uses were reported, including privacy leakage detection (3 respondents), source code viewing (3 respondents), study (2 respondents), library logic understanding (1 respondent), and tool refinement (1 respondent).
Regarding the usage experience of Java decompilers, 76\% respondents use decompilers on a weekly or monthly basis This frequent usage suggests that most respondents are well-acquainted with these Java decompilers, which in turn implies that their evaluations of the Java decompilers are likely to be relatively objective. 

Our survey also suggests that 24\% of our respondents think that Java decompilation is extremely important (10 scores), with an average score of 7.7.
% Figure~\autoref{fig:survey2}
\autoref{fig:survey2} further illustrates the distribution of respondents' evaluation on the correctness and understandability of Java decompilation. % 32/33 = 0.9696  31/33 = 0.9393
In detail, 97\% of respondents assigned a score over 5 to the importance of decompilation correctness, while the corresponding figure for the importance of decompilation understandability is 94\%. Their average scores are 8.5 and 8.3, respectively. Notably, a score of 1 denotes ``not important at all," and a score of 10 denotes ``very important," with a score of 5.5 being the midpoint. Although 6\% (2 out of 33) of respondents rated the importance of decompilation understandability below 5.5, 33\% (11 out of 33) considered it very important (score 10), a proportion even higher than that for decompilation correctness. It is to be expected that stakeholders of Java decompilers value both decompilation correctness and understandability. Particularly noteworthy is that these statistics suggest that developers and researchers in the wild regard decompilation understandability as important as its correctness.

% frequency of understandability issues
% 3 3 1 3 6 0 6 8 2 1 = 33
% [1, 2, 3, 4, 5, 6, 7, 8, 9, 10]
% [3, 3, 1, 3, 6, 0, 6, 8, 2, 1]
Similar trends have also been observed in the frequency of decompilation failures and understandability issues encountered by respondents who have used decompilers before. The average scores for the frequency of encountered decompilation failures and decompilation understandability issues are 4.8 and 5.7, indicating that understandability issues are more commonly encountered in daily Java decompiler usage. 
A detailed investigation reveals polarized opinions on the frequency of understandability issues: though 30\% respondents (10 out of 33) rated the frequency of understandability issues as low (score 4 or lower), 52\% regraded them as frequent (score 7 or higher), and 24\% assigned a high score of 8, the most frequently-selected score among all options. % 1-4:10  5:6  7-10:17  10/33=0.303 17/33=0.5158/3 8/33=0.242
While it is possible that a number of understandability issues arise from the inherent complexity of the source code, our empirical findings addressing RQ2 indicate that a significant proportion of these issues are caused by the implementation of Java decompilers themselves. This part is further demonstrated in Section~\ref{sec:RQ2}.

\begin{result}
Java program developers and researchers in the wild regard the understandability of Java decompilation as important as its correctness, and decompilation understandability issues are even more commonly encountered than decompilation failures.
\end{result}

Decompilation correctness has garnered considerable attention and has been extensively studied~\cite{harrand2020java, mauthe2021large, lu2024}. Unfortunately, no existing studies have ever focused on the understandability of decompilation. Our work aims to fill this gap by conducting the first comprehensive study on decompilation understandability. We investigate the typical causes of understandability issues, explore the potential for developing metrics to effectively assess understandability, and pave the way for future researches in this area.

\section{Understandability Metrics}\label{sec:background}

In this section, we introduce Cognitive Complexity and perplexity, which stand for rule-based and language-model-based metrics for code understandability, respectively. %and n-gram language models.

\subsection{Cognitive Complexity}

Cognitive Complexity~\cite{campbell2018cognitive} is introduced as a novel approach for assessing code understandability, aiming to address the limitations of Cyclomatic Complexity~\cite{mccabe1976complexity}. Cognitive Complexity is evaluated based on three basic rules: First, ignore structures that enable concise representation of multiple statements in a readable manner. Second, apply an increment for each interruption in the linear flow of the code. Various statements including $if$, $else$, $for$, $do$, $while$, $catch$, $switch$, $goto$, $break$, $continue$, as well as ternary operators and sequences of different logical operators contribute to the increment. Third, apply an increment when flow-breaking structures are nested. This rule entails that it takes into account the nesting depth when assessing the overall complexity of code. 

%  float, floatplacement=H, escapechar=\^,
\textbf{}\begin{lstlisting}[style=Java, caption={An example of Cognitive Complexity.}, label={lst:cog-example}, belowskip=-5\baselineskip, columns=flexible, basicstyle=\scriptsize]
public boolean containsDigitOrLetter(Character[] input) {
  for (Character c : input) {                               //+1
    if (c != null                                           //+2 (nesting=1)
       && (Character.isDigit(c) || Character.isLetter(c)))  //+2
      return true;
  }
  return false;
}                                    // Cognitive Complexity = 5
\end{lstlisting}

% The example is removed temporarily
An illustrative example of the calculation process of  Cognitive Complexity is provided in \autoref{lst:cog-example}. In line 2, the Cognitive Complexity score increases by 1 when encountering $for$. Simultaneously, the nesting score increases by 1 due to the introduction of a nested structure. line 3 contributes to the Cognitive Complexity score with an increase of 2, which results from the presence of a $if$ statement, and the nesting score, currently at 1. Finally, in line 4, the Cognitive Complexity score increases by 2, which reflects the cumulative impact of various logical operators.

Cognitive Complexity is validated as a measure of source code understandability in a study~\cite{munoz2020empirical}, which has demonstrated a positive correlation between Cognitive Complexity and both developers' comprehension time and their subjective ratings of code understandability. 

% Building upon these findings, 
% . This research represents a promising endeavor to leverage a well-established metric in a novel context
We seek to extend the utility of Cognitive Complexity by applying it to the evaluation of decompiled code understandability, with the goal of shedding light on assessing code understandability in the context of decompilation.

\subsection{Perplexity}
% assess the coherence of a given sequence of words from the perspective of models. 
Perplexity~\cite{manning1999foundations} of language models is the exponential form of cross-entropy. It is utilized to evaluate the performance of language models, and it can also be employed to quantify how surprising a sentence appears in the context of the model's probability distribution. A higher perplexity score indicates that the sequence of words may deviate from what the model typically expects, implying potential issues. 
For a sequence $S$ with $n$ words, perplexity is defined as follows:
$$
\textit{Perplexity}(S) = \exp \left\{ -\frac{1}{n} \sum_i^n \log P(w_i|w_{<i})  \right\}
$$
in which $w_i$ is the i-th words of $S$, and $P(w_i|w_{<i})$ is the conditional probability of the i-th word given preceding words.

As for language models, we employ n-gram language models due to their extensive prevalence in software engineering tasks. Hindle et al.~\cite{hindle2012naturalness} have demonstrated that code exhibits repetitive patterns and possesses predictable statistical properties that can be effectively captured by n-gram language models. Subsequently, n-gram language models have been widely used in various software engineering tasks including bug detection, code completion and so on~\cite{allamanis2013mining, wang2016bugram, raychev2014code, tu2014localness, allamanis2018survey}.
% Allamanis et al.\cite{allamanis2013mining} have suggested that n-gram log probability can be used as code complexity measures. 
In n-gram language models, preceding n-1 words are considered as the context for predicting the probability of the next word. When n is set to 5, the conditional probability is
$$
P(w_n|w_1,w_2,...,w_{n-1}) = \prod_{i=1}^{n}P(w_i|w_{i-1}w_{i-2}w_{i-3}w_{i-4})
$$
in which $P(w_i|w_{i-1}w_{i-2}w_{i-3}w_{i-4})$ is the conditional probability of the i-th word given the preceding four words.
% This finite context window reflects the Markovian assumption that the current word's occurrence is dependent only on a fixed, limited span of prior words within the sequence. This simplifying assumption enables the modeling of complex language patterns while reducing computational complexity.

Our research endeavors to determine whether perplexity of n-gram language models correlates with the understandability of decompiled code. % We aim to reveal valuable insights into the relationship between language model-based metrics and the understandability of decompiled code.

\begin{figure}[tbp]
\centerline{\includegraphics[width=0.48\textwidth]{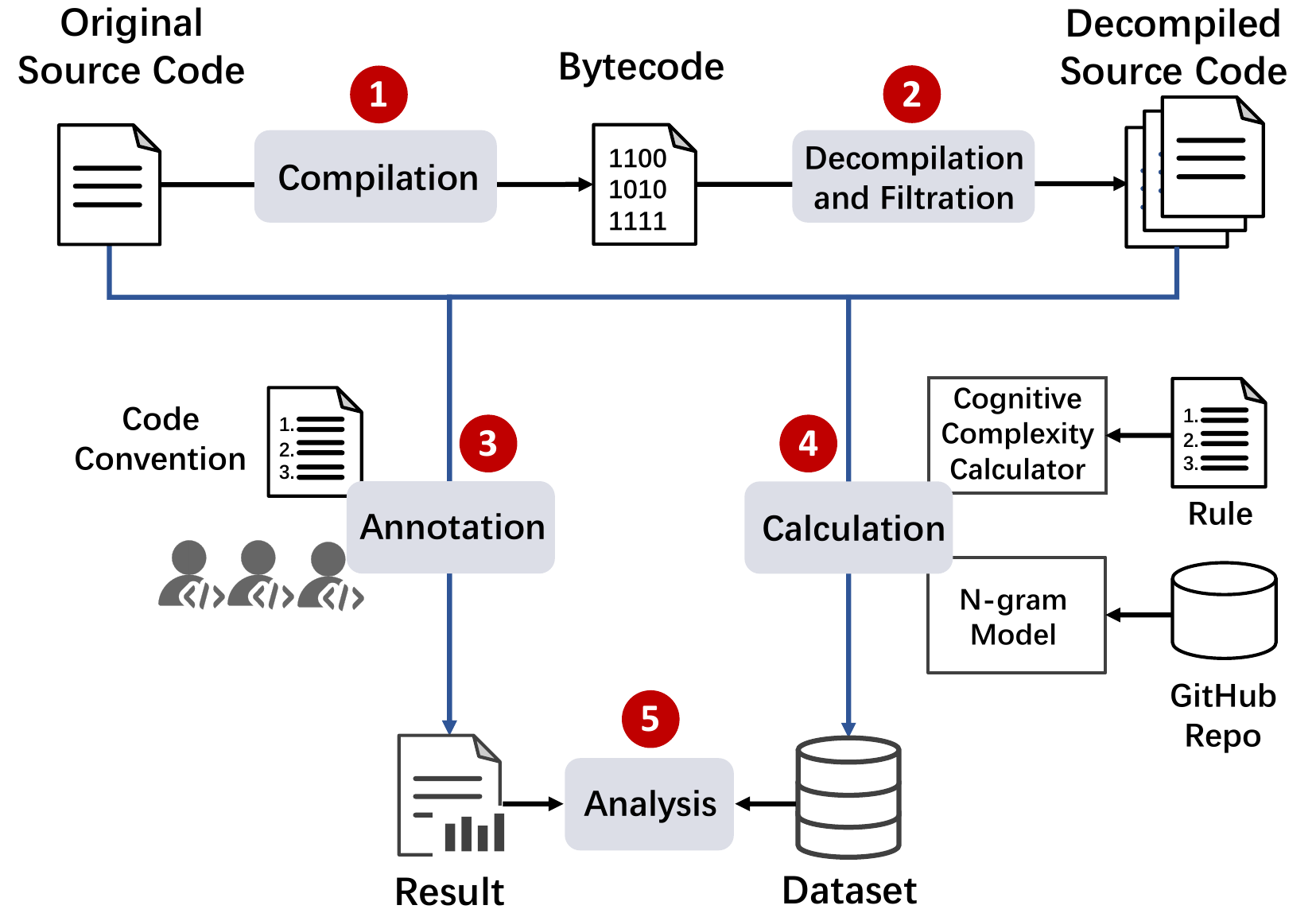}}
\caption{Framework including compilation, decompilation, filtration, calculation, annotation and analysis.}
\label{fig:framework}
\end{figure}

\section{Experimental Methodology}\label{sec:methodology}

% Subsequently, 
To provide a micro picture of the understandability of Java decompilation, a series of three experiments has been conducted to investigate into the understandability of Java decompilation in a micro manner. In detail, through these experiments, we delve deeper into the understandability disparity between the source code from real-world Java projects and the corresponding decompiled code generated by real-world Java decompilers quantitatively. Since no existing benchmarks has ever been proposed for the Java decompilation understandability, the first experiment tries to give an as objective and impartial as possible assessment on the understandability, with understandability rules and manual assistance. 
This experiment serves to demonstrate the inherent diversity in the understandability between the decompiled code and the source code, thereby establishing a ground truth for the following understandability assessment. 
The following two experiments, therefore, are designed to investigate the effectiveness of different kinds of metrics in recognizing understandability disparities, which address our RQ3 and RQ4. In these two experiments, Cognitive Complexity and Perplexity, standing for rule based metrics and language-model-based metrics respectively, are selected for the understandability assessment. %The first set focu The first set focuses on rule-based understandability metrics, aiming to evaluate their efficacy in assessing the understandability of both decompiled code and their corresponding source code, thereby directly addressing RQ3. The second set of experiments centers on language-model-based metrics, seeking to ascertain their capacity to assess the understandability of both decompiled code and their corresponding source code, directly addressing RQ4. Through this comprehensive methodological approach, this study endeavors to provide nuanced insights into the multifaceted dimensions of code understandability within the realm of Java decompilation.

% In this section, we introduce our methodology and provide an overview of the framework for the analysis of code understandability within the context of Java decompilation. 

% To answer RQ1, we conduct a survey on Java Decompilation to better understand opinions from the community via questionnaire. 

\subsection{Setup for Understandability Assessment Experiments}

To set up the following three experiments regarding understandability between the decompiled code and source code from real-world Java decompilers and projects, we propose an experiment framework as illustrated in~\autoref{fig:framework}. 

\subsubsection{Experimental Protocol}
In this framework, the three experiments have been conducted with the following steps:
\begin{enumerate*}
\item[(1)] Compilation: First, the source code extracted from real-world Java projects, denoted as the original source code, is compiled into bytecode with \project{javac} of version 1.8.0351.
\item[(2)] Decompilation and Filtration: The bytecode is then decompiled using various decompilers, and the corresponding output is referred to as the decompiled code. 
Since this study focus on decompilation understandability, we would like to exclude the failed decompilation cases in our statistics, because though failures in decompilation, i.e., exceptions, syntactic errors and inconsistent semantics, may greatly influence the decompilation understandability, however, they are usually due to bugs in the decompilers and should not be the main focus of our paper.
%As we focus on files that maintain both syntactic correctness and semantic equivalence, we filter them out by recompiling and testing.
\item[(3)] Annotation: To provide a ground truth against which two metrics can be evaluated, an annotation of the understandability of the decompiled code has been conducted. During this process, both convention for code understandability and manual assistance have been employed to ensure quality of the annotation and mitigate the potential bias in this assessment. 
\item[(4)] Metric Calculation: To evaluate Cognitive Complexity and perplexity, we compute scores of them for both decompiled code and original source code.
\item[(5)] Result Analysis: To comprehensively validate the effectiveness of the metrics in assessing the code understandability in the context of decompilation, we analyze outcomes derived from our annotated dataset and two metrics. % and their subsequent evaluation against our annotated dataset.
\end{enumerate*}

\subsubsection{Experimental Objects}
As for experimental objects, we focus on four extensively utilized Java decompilers following the prior research~\cite{lu2024}, namely \project{CFR}~\cite{cfrgithub}, \project{Fernflower}~\cite{fernflowergithub}, \project{JD-GUI}~\cite{jdGUIgithub}, and \project{Jadx}~\cite{jadxgithub}, which are the four most popular Java decompilers as detailed in Section~\ref{sec:survey}. However, we eventually excluded \project{JD-GUI} from our experiments due to its inactive development status, evidenced by its last update in 2019. Our selection of experiment objects also aligns with a related work targeting Java decompilers~\cite{lu2024}.
An overview of the Java decompilers' characteristics is provided in \autoref{tab:decompilers}. Notably, all of them have gained over 1,000 stars in GitHub, proving their popularity. It is also worth noting that \project{Fernflower}, has been widely used within IntelliJ IDEA~\cite{intellij}, one of the most popular and widely used IDEs focused on Java language, for decompiling Java classes since version 14.1.

\begin{table}[htbp]\footnotesize
\caption{Characteristics of three decompilers.}
% \scriptsize
    \setlength{\tabcolsep}{0.35em}
\centering
    \begin{tabular}{|l|c|c|c|c|}
		\hline
		\textsc{\textbf{Decompiler}} & \textsc{\textbf{Version}} & \textsc{\textbf{ \#Commits}} & \textsc{\textbf{\#Stars}} & \textsc{\textbf{\#LOC}}\\
		\hline
            \project{CFR}	     & 0.152 & 1878 & 1.7K & 66173\\ \hline
            \project{Fernflower} & -     & 637  & 3K   & 57264\\ \hline
            \project{Jadx}	     & 1.4.5 & 1861 & 36K  & 69214\\
		\hline
	\end{tabular}
	\label{tab:decompilers}
\end{table}

\begin{table}[htbp] \footnotesize
\caption{An overview of files after compilation, decompilation and filtration (\textsc{\textbf{ORI}}: Original. \textsc{\textbf{REC}}: Recompilable. \textsc{\textbf{PAT}}: Able to pass tests.).}
% \scriptsize
\centering
    \setlength{\tabcolsep}{0.45em}
    \begin{tabular}{|l|c|c|c|}
		\hline % \tnote{1}
        \textsc{\textbf{Decompiler}} & \textsc{\textbf{\#ORI}} & \textsc{\textbf{\#REC}} & % Recompilable
        \textsc{\textbf{\#PAT}} \\ %PassTest Filterd & \textsc{\textbf{\#LOC}}\\
		\hline
        \project{CFR} & $2041$ & $1739$ & $1730$ \\
        \hline
        \project{Fernflower} & $2041$ & $1387$ & $1378$ \\
        \hline
        \project{Jadx}	& $2041$ & $1374$ & $1346$ \\
        \hline
        \textsc{\textbf{Intersection}} & \textbf{$2041$} & \textbf{$1175$}  & \textbf{$1150$} \\
        \hline
    \end{tabular}
    
    % \begin{tablenotes}
    %   \item[1] \textsc{\textbf{ORI}}: Original. \textsc{\textbf{REC}}: Recompilable.  \textsc{\textbf{PAT}}: Able to pass tests.
    % \end{tablenotes}
    
    \label{tab:decompile_class_overview}
\end{table}

\subsubsection{Experimental Subjects}
To obtain a collection of real-world Java projects for the assessment, we have employed the same set of projects as previous works~\cite{harrand2020java,pawlak2016spoon}, including 2041 Java files from 14 projects. We leverage the tool proposed in the previous study~\cite{harrand2020java} to process compilation, decompilation and filtration. \autoref{tab:decompile_class_overview} provides an overview of the outcomes after these processes, including original files (ORI),  recompilable files (REC) and files able to pass tests (PAT).

% Because files without positive scores are relatively simple   may not yield significant research insights
% After calculation, we retain files that exhibit positive values for both of the two metrics. This selection process excludes relatively simple files, including straightforward enumerations, interfaces and classes primarily comprised of getter and setter methods.
% review: Files that exhibit a Cognitive Complexity score of 0 which have been excluded after calculation.

% add 0512 for SANER24  It is unclear from my side at which granularity level these differences have been observed. Specifically, when a source code was decompiled, the only differences can be observed in how the functions are implemented. Without this information, it is hard for me to understand the soundness of this study.  

In alignment with prior research, we conducted our experiment at the file granularity level, as files serve as the fundamental entities in the process of decompilation. To ensure a more focused investigation into decompilation challenges and the relevance of our analysis, we excluded simple files such as enumerations, interfaces, and classes that consist solely of straightforward getter and setter methods. An overview of these selected files is provided in \autoref{tab:projects_overview}, specifying the projects from which they originate. After this step, we obtain a subset comprising 429 sets of files, which are expected to offer valuable insights into the understandability of decompiled code.

Note that these projects utilize \project{Apache Maven} as management tools, with the majority of them employing the default debug option, configured as \textit{true}. Consequently, during decompilation, debugging information including variable names can be recovered. However, in projects that opt to remove debugging information, variable names become irretrievably lost. Therefore, our assessment of decompiled code understandability does not take variable names into consideration.

\subsubsection{Experimental Metrics}
To answer RQ2, we referred to coding conventions from three world-famous companies, namely, Oracle~\cite{oracle2000code}, Google~\cite{googleguide}, and Alibaba~\cite{aliguide}---due to the absence of universally accepted standards for code understandability. These conventions serve as  oracles for our assessment of code understandability.
However, it is important to note that these conventions are guidelines with recommendations on various aspects of software development. Given the considerable diversity in the structures of real-world code snippets, automating the assessment process remains challenging. Therefore, two graduate students were engaged to annotate the relative understandability of these files. Both students have completed coursework in software engineering and have over five years of experience in Java programming. 
To prevent potential bias as possible as we can, it took 15 hours for both students to be familiar with the three coding conventions before annotation. 
This familiarity with the coding conventions provides a foundation for the subsequent annotation process. % including Java coding conventions of Oracle~\cite{oracle2000code}, Google Java Style Guide~\cite{googleguide} and Alibaba Java Coding Guidelines~\cite{aliguide}.   ~\cite{oracle2000code,googleguide,aliguide}.

% Annotation of Understandability: 
During the annotation process, both students were required to individually assess the relative understandability of the decompiled files compared to the corresponding original source files. 
For a given pair of decompile file and source file, relative understandability is classified into three categories: \textbf{\textit{Less}}, indicating that the decompiled file is significantly less understandable than the source file; \textbf{\textit{Equi}}, signifying that the understandability of the two files is relatively comparable; and \textbf{\textit{More}}, indicating that the decompiled code is significantly more understandable. Noteworthy, supporting evidence from the coding conventions had to be provided by both students for reference.
A classification could only be finalized when both students agreed. In cases of disagreement, a third student, also familiar with the coding conventions and experienced in Java programming, was tasked with making the final decision.
% Final Annotated Dataset:
Finally, we established a final annotated dataset that represents the relative understandability of decompiled files compared to the original counterparts.
% along with the associated reasons.

For the metric calculation, an open-source tool~\cite{cong2023} was employed to measure Cognitive Complexity of code snippets. As for perplexity, a 5-gram language model with highly popular Java projects collected from \textit{GitHub} has been trained. In order to ensure the high quality of our training dataset, we curated 40 Java projects, each of which has gained more than 10,000 stars and 5,000 forks on \textit{GitHub}. This meticulous selection encompasses a total of 107,696 Java files, serving as the foundation for the training of our model.

\begin{table}[htbp]\footnotesize
\caption{An overview of selected files (\textsc{\textbf{ORI}}: Original. \textsc{\textbf{REC}}: Recompilable.  \textsc{\textbf{PAT}}: Able to pass tests. \textsc{\textbf{EVA}}: To be evaluated.).}
\centering
    \setlength{\tabcolsep}{0.35em}
    \begin{tabular}{|l|c|c|c|c|c|c|}
\hline
% \toprule
% \multirow{2}{*}{ \textsc{\textbf{Project}} }  & \multicolumn{2}{c|}{ \textsc{\textbf{Original}} }  & \textsc{\textbf{REC}} & \textsc{\textbf{PAT}} & \multicolumn{2}{c|}{ \textsc{\textbf{Evaluate}} }\\
% \cline{2-7}
% & \textsc{\textbf{\#}} & \textsc{\textbf{LOC}} & \textsc{\textbf{\#}}    & \textsc{\textbf{\#}}    & \textsc{\textbf{\#}} & \textsc{\textbf{LOC}} \\
\textsc{\textbf{Project}} & \textsc{\textbf{\#ORI}} & \textsc{\textbf{\#REC}} & \textsc{\textbf{\#PAT}} & \textsc{\textbf{\#EVA}} & \textsc{\textbf{\#ORI\textsubscript{LOC}}} & \textsc{\textbf{\#EVA\textsubscript{LOC}}} \\\hline
% \midrule 
        Bukkit                  & $642$ & $479$ & $479$ & $100$ & $60800$ & $11453$ \\ \hline
        Codec                   & $59$  & $44$  & $44$  & $26$  & $15087$ & $7301$  \\ \hline
        Collections             & $301$ & $54$  & $53$  & $32$  & $62077$ & $3437$  \\ \hline
        Imaging                 & $329$ & $186$ & $185$ & $86$  & $47396$ & $8099$  \\ \hline
        Lang                    & $154$ & $55$  & $51$  & $28$  & $79509$ & $6983$  \\ \hline
        DiskLruCache            & $3$   & $1$   & $0$   & $0$   & $1206$  & $0$     \\ \hline
        JavaPoet                & $2$   & $1$   & $0$   & $0$   & $934$   & $0$     \\ \hline
        Joda time               & $165$ & $114$ & $104$ & $71$  & $70027$ & $18011$ \\ \hline
        Jsoup                   & $54$  & $19$  & $18$  & $11$  & $14801$ & $1544$  \\ \hline
        JUnit4                  & $195$ & $107$ & $102$ & $36$  & $17167$ & $3391$  \\ \hline
        Mimecraft               & $4$   & $2$   & $2$   & $2$   & $523$   & $304$   \\ \hline
        Scribe Java             & $89$  & $80$  & $80$  & $25$  & $4294$  & $1411$  \\ \hline
        Spark                   & $34$  & $27$  & $26$  & $9$   & $4089$  & $929$   \\ \hline
        DcTest                  & $10$  & $7$   & $6$   & $3$   & $211$   & $102$   \\
\hline
% \midrule
        \textsc{\textbf{Total}} & \textbf{$2041$} & \textbf{$1175$} & \textbf{$1150$} & \textbf{$429$}  & \textbf{$378121$}  & \textbf{$62965$} \\
\hline
% \bottomrule
	\end{tabular}
    % \begin{tablenotes}
    %   \item[1] \textsc{\textbf{ORI}}: Original. \textsc{\textbf{REC}}: Recompilable.  \textsc{\textbf{PAT}}: Able to pass tests. \textsc{\textbf{EVA}}: To be evaluated.
    % \end{tablenotes}
    
    \label{tab:projects_overview}
\end{table}

\section{Experimental Results}\label{sec:results}

\begin{table}[htbp]\footnotesize % \footnotesize
\caption{Details of the relative understandability of the annotated dateset (Each cell denoted as $a/b/c$ presents that under the specific Java project and decompiler, there are $a$, $b$, and $c$ decompiled files demonstrating \textbf{\textit{Less}}, \textbf{\textit{Equi}}, and \textbf{\textit{More}} understandability relative to their respective source code.).}
\centering
    \setlength{\tabcolsep}{0.3em}
    \begin{tabular}{|l|c|c|c|c|c|}
		\hline
		\textsc{\textbf{Project}} & \textsc{\textbf{\#File}} & \textsc{\textbf{CFR}} & \textsc{\textbf{Fern}} & \textsc{\textbf{Jadx}} & \textsc{\textbf{Total}}\\
		\hline
        Bukkit	            & 300 & 19/63/19 & 24/70/6 & 17/76/7 & 59/209/32   \\ \hline
        Codec               & 78  & 12/14/0  & 14/12/0 & 8/18/0  & 34/44/0     \\ \hline
        Collections	        & 96  & 3/29/0   & 1/31/0  & 1/31/0  & 5/91/0      \\ \hline
        Imaging	            & 258 & 34/51/1  & 33/53/0 & 13/72/1 & 80/176/2    \\ \hline
        Lang                & 84  & 13/15/0  & 11/17/0 & 6/22/0  & 30/54/0     \\ \hline
        Joda time 	        & 213 & 18/53/0  & 17/54/0 & 9/62/0  & 44/169/0    \\ \hline
        Jsoup               & 33  & 0/2/9    & 0/3/8   & 0/3/8   & 0/8/25      \\ \hline
        Junit4              & 108 & 5/29/2   & 4/30/2  & 2/32/2  & 11/91/6     \\ \hline
        Mimecraft           & 6   & 0/2/0    & 0/2/0   & 1/1/0   & 1/5/0       \\ \hline
        Scribe Java         & 75  & 2/19/4   & 1/20/4  & 0/21/4  & 3/60/12     \\ \hline
        Spark   	        & 27  & 1/8/0    & 0/9/0   & 0/9/0   & 1/26/0      \\ \hline
        DcTest              & 9   & 0/1/2    & 1/0/2   & 1/0/2   & 2/1/6       \\ 
        \hline
    \textsc{\textbf{Total}} &1287&106/286/37&106/301/22&58/347/24 & 270/934/83  \\
        \hline
	\end{tabular}
	\label{tab:dataset_detail}
\end{table}

\subsection{RQ2: how can code understandability vary between decompiled code and source code?} 
\label{sec:RQ2}

% group by decomplers 
The annotated result of 1287 files is presented in \autoref{tab:dataset_detail} by Java projects and decompilers.  %Among the total 1287 files, 270, 934, and 83 decompiled files exhibit \textbf{\textit{Less}}, \textbf{\textit{Equi}}, and \textbf{\textit{More}} understandability compared to their corresponding source code, respectively.
Evidently, a predominant 93.6\% ((934+270)/1287) of the decompiled files, are annotated as either \textbf{\textit{Equi}} or \textbf{\textit{Less}} understandable in comparison to their original source code files. This observation is consistent with logical expectations, given that decompiled files share identical semantics (bytecode) with their corresponding source code, where the semantics greatly influence the understandability of both the decompiled and source code. However, it is important to note that the source code is typically maintained and optimized with human-intelligence, and naturally should have higher readability and understandability than code automatically generated by Java decompilers. This may explain the observed 21.0\% decompiled files exhibiting \textbf{\textit{Less}} understandability. 

% \begin{table}[htbp]\scriptsize
% \caption{An overview of relative understandability of the decompiled files when compared to their corresponding ones.}
% \centering
%     \setlength{\tabcolsep}{0.35em}
%     \begin{tabular}{l|cccc}
% 		\hline
% 		 & \textsc{\textbf{CFR}} & \textsc{\textbf{Fernflower}} & \textsc{\textbf{Jadx}} & \textsc{\textbf{All}} \\ \hline
%         \#Less 	      & 1  & 2 & 3 & 270 \\
%         \#Equi          & 4  & 5 & 6 & 935 \\
%         \#More 	      & 7  & 8 & 9 & $82(6\%)$   \\ \hline
% \textsc{\textbf{Total}} & $429(100\%) $ & $429(100\%)$&$429(100\%)$& $1287(100\%)$ \\
%         \hline
% 	\end{tabular}
% 	\label{tab:dataset_annotated}
% % \vspace{-0.5cm}
% \end{table}

% \autoref{fig:dataset_annotated} provides a higher-level profile of the understandabilty of the decompiled code and their corresponding source code.
% Obviously, 21\% of files have significant lower understandability after decompilation among the three projects. 

\begin{figure}[tbp]
\centerline{\includegraphics[width=0.4\textwidth]{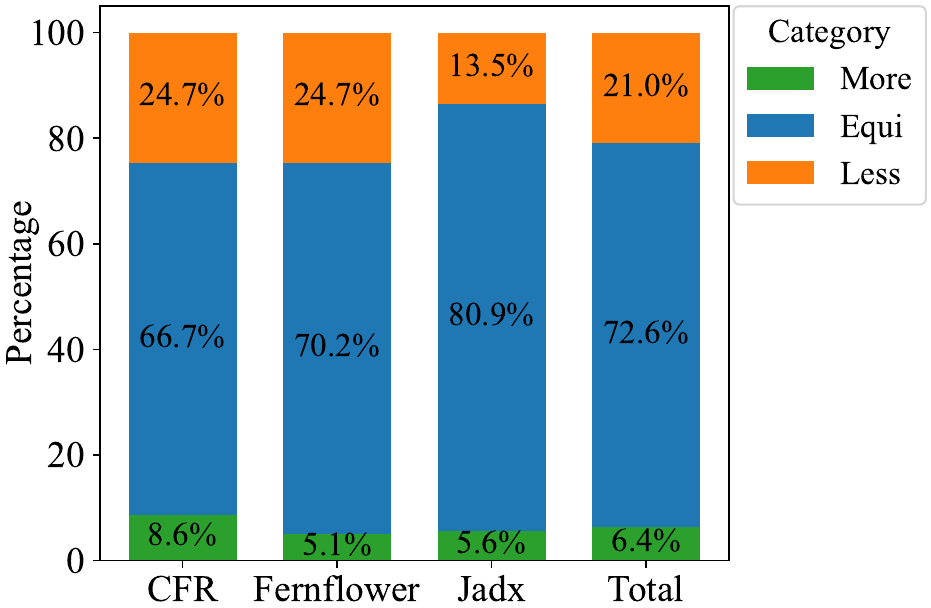}}
\caption{An overview of relative understandability of the decompiled files when compared to corresponding original ones.}
\label{fig:dataset_annotated}
\end{figure}

While this proportion may be relatively modest, such cases of \textbf{\textit{Less}} can be commonly observed in eleven out of all twelve Java projects (the exception is project \project{Jsoup} containing only 33 files) and the percentage of all three decompilers as shown in \autoref{fig:dataset_annotated}. This observation indicates that, despite decompilers' ability to generate code that is syntactically and semantically accurate, a noteworthy subset of the decompiled code manifests a higher complexity than that of the source code, which, however, may impede the intended objective of program comprehension in decompilation.
Meanwhile, these files also underscore the inherent challenges in generating well-structured code in decompilation, thereby indicating potential areas for enhancement in the code generation strategies employed by Java decompilers. 
%  This issue represents a noteworthy concern, given its potential implications for the effectiveness of existing decompilers.

For the remaining 6.4\% decompiled files, they are identified to have higher understandability relative to their original source code files. This data implies the potential of Java decompilers to generate code snippets that exhibit superior structural coherence and understandability, even in comparison to source code written by developers. %It is a surprising phenomenon, since these cases indicate that the generation strategies used in Java decompilers enhance the code understandability even compared with source code, instead. An interesting finding is that these cases are project-specific, and are mainly concentrated in projects \project{Bukkit}, \project{Jsoup}, and \project{Scribe Java}, etc.

Furthermore, an in-depth analysis of the differences between the understandability of the decompiled code and the original source code has been conducted. This detailed analysis is presented in Section~\ref{sec:casestudy}, which investigates the root causes of the understandability disparities in terms of code patterns.

\begin{result}
A noteworthy proportion of code snippets decompiled by current Java decompilers exhibits significant lower or even higher understandability relative to their respective source code.
\end{result}

\subsection{RQ3: how can rule-based understandability metrics assess the understandability of both decompiled code and their corresponding source code?} \label{subsec:rq3}

\begin{figure*}[htbp]
\includegraphics[width=0.33\textwidth]{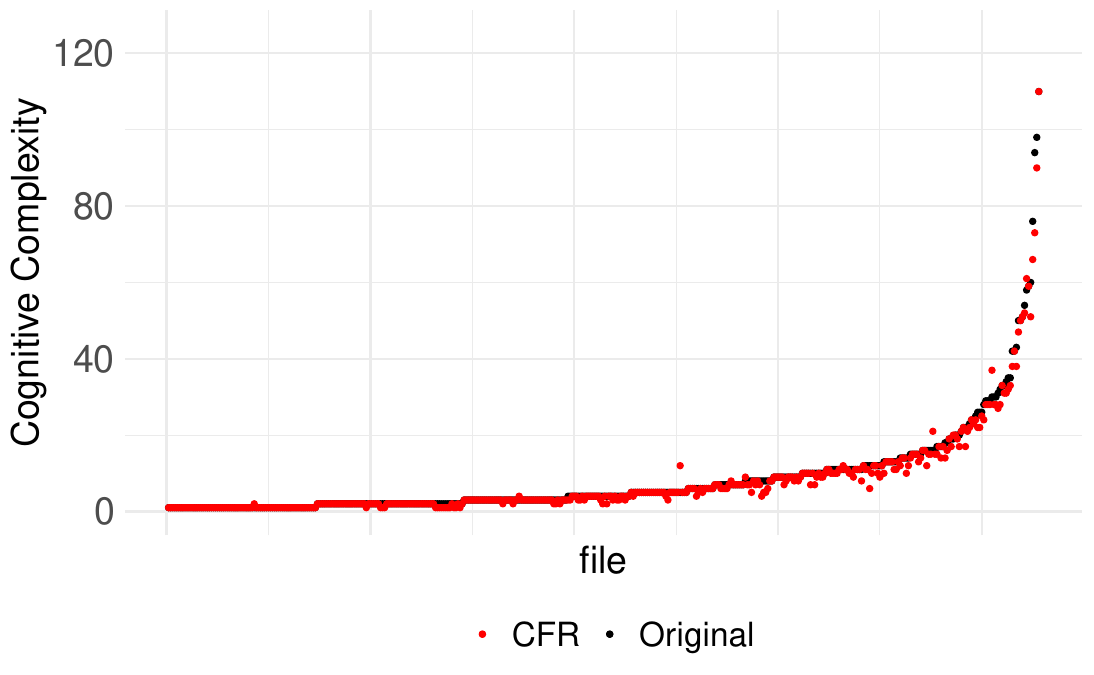}
\includegraphics[width=0.33\textwidth]{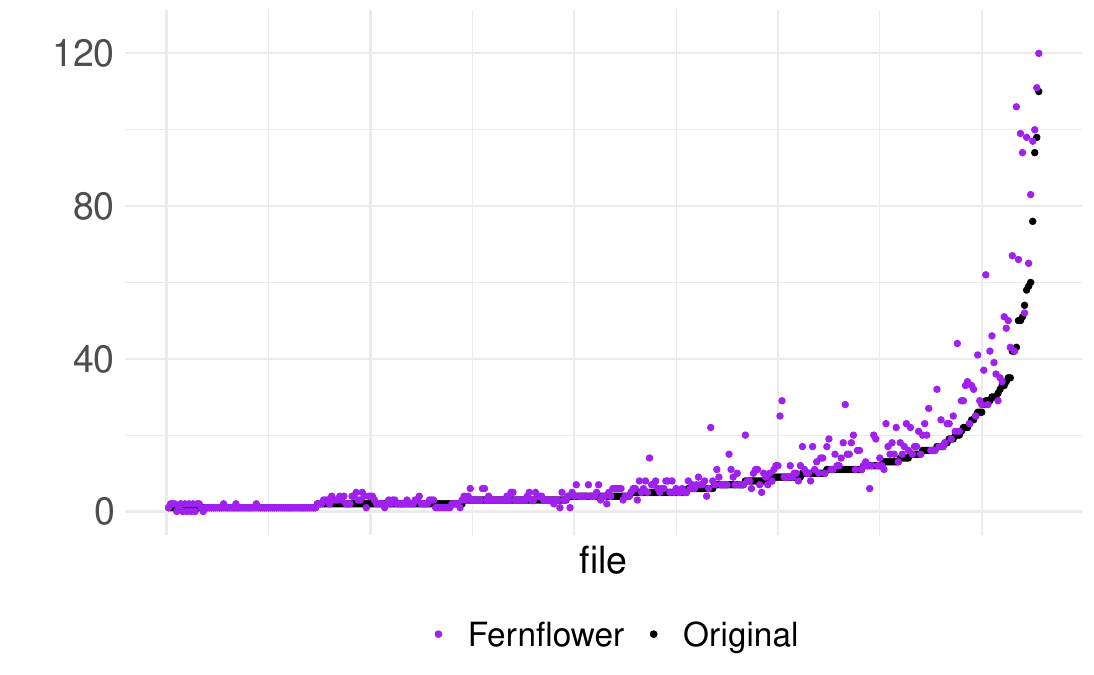}
\includegraphics[width=0.33\textwidth]{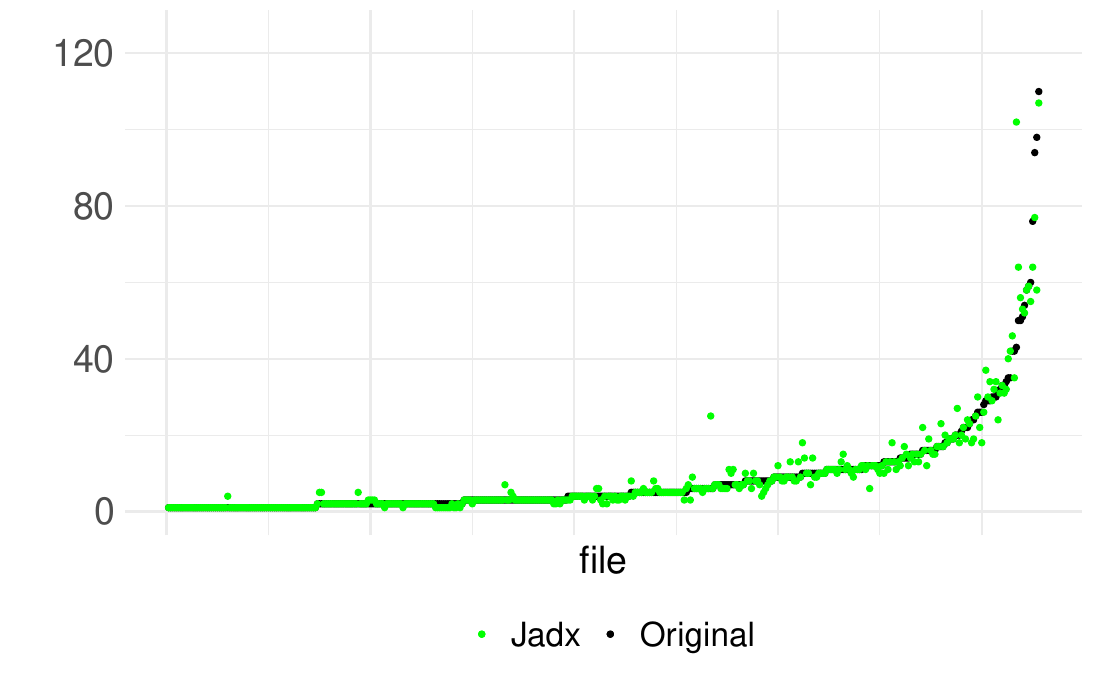}
\caption{Scatter plots of Cognitive Complexity scores of original files and corresponding decompiled files.} \label{fig:scatter_cc}
\end{figure*}

To answer RQ3, \autoref{fig:scatter_cc} presents Cognitive Complexity scores of the original files and corresponding decompiled files with an outlier removed. The files in the \autoref{fig:scatter_cc} are organized in ascending order based on their respective Cognitive Complexity of original files. The only outlier pertains to a source file in the project \project{Lang}, whose Cognitive Complexity is 138. The associated decompiled code, generated by \project{CFR}, \project{FernFlower}, and \project{Jadx}, manifests Cognitive Complexity scores of 150, 197, and 177, respectively.
The Cognitive Complexity scores for the remaining source files exhibit a spectrum ranging from 1 to 110, with an average value of 8.5. With decompiler \project{CFR}, \project{Fernflower}, and \project{Jadx}, the scores of respective decompiled files range from 1 to 110, 0 to 120, and 1 to 107, yielding an average of 8.0, 11.0, and 8.6.
%Among the twelve projects, the Cognitive Complexity metrics for their source code files exhibit a spectrum ranging from 1 to 138, with an average value of 8.8. With decompiler \project{CFR}, \project{Fernflower}, and \project{Jadx}, the Cognitive Complexity of respective decompiled files ranges from 1 to 150, 0 to 197, and 1 to 177, yielding an average of 8.3, 11.4, and 9.0.

Notably, the scores for files generated by \project{CFR} 
and \project{Jadx} closely resemble those of the original files. In contrast, files produced by \project{Fernflower} exhibit a significant increase when compared to their original counterparts because \project{Fernflower} tends to generate explicit $else$ statements and braces, resulting in more nested code structures and higher scores. % because \project{Fernflower} tends to generate more deeply nested code structures. %This phenomenon can be attributed to the tendency of \project{Fernflower} to produce code structures with greater levels of nesting.

% These insights shed light on the challenges associated with using Cognitive Complexity as a measure of decompiled code understandability in the context of different decompilers and their unique behaviors.

While two out of the three Java decompilers turn out to generate code with Cognitive Complexity levels comparable to that of the source code, whether these code pieces exhibit similar levels of understandability warrants further exploration.
Therefore, we compute the confusion matrix of the relative understandability indicated by Cognitive Complexity with reference to the annotated dataset. In order to convert Cognitive Complexity into three categories of relative understandability, we consider files with similar Cognitive Complexity scores to the original equivalently understandable. 
Specifically, we employ the following formula:
\[
U(x,ori) = 
\begin{cases}
    Less, & \text{if } x > ori+t \\
    Equi, & \text{if } ori-t \leq x \leq ori+t \\
    More, & \text{if } x < ori-t
\end{cases}
\]
Here $x$ and $ori$ represents the Cognitive Complexity score of the decompiled and the original file, respectively, where $t$ represents the threshold denoting the maximum allowable disparity between the decompiled and original file. For the value of $t$, we tried integer values ranging from 0 to 10 to optimize the overall performance.
Given the imbalanced nature of the dataset, we choose to maximize the macro F1-score, which is the mean of F1-scores of all classes. After comprehensive testing, we determined that setting t to 3 resulted in the highest macro F1-score, achieving a value of 0.47. This specific threshold balances precision and recall across different classes in the dataset, thereby providing the best overall performance.

% review: It is not clear why the authors selected t = 3 for answering RQ2. How did they find that this threshold allows them to achieve the highest F1-score? What threshold values did they test?

% This selection is underpinned by the observed robustness of F1-scores associated with this threshold within the context of this study. Further elaboration on the rationale behind this choice will be provided subsequently.

% \input{tables/confusion_matrix_CC_ori}
\begin{table}[htbp]
    \caption{Confusion Matrix and metrics of relative understandability indicated by Cognitive Complexity.} \label{tab:conf_matrix_CC_ori_all}
    % \centering
    \begin{subtable}[htbp]{0.47\linewidth} \footnotesize
        % \centering
        \caption{Confusion Matrix}
    \setlength{\tabcolsep}{0.35em}
            \begin{tabular}{|c|c|c|c|}
            \hline
                \diagbox{\textbf{Predict}}{\textbf{Actual}}  & Less & Equi & More \\\hline
                Less & 79 & 19 & 4 \\
                \hline
                Equi & 181 & 901 & 72 \\
                \hline
                More & 10 & 14 & 7 \\
                \hline
            \end{tabular}
        \label{tab:conf_matrix_CC_ori}
    \end{subtable}
    \hfill
    \vspace{0.2cm}
    \begin{subtable}[htbp]{0.49\linewidth} \footnotesize
        \centering
        \caption{Metrics}
    \setlength{\tabcolsep}{0.35em}
        \begin{tabular}{|c|c|c|c|}
        \hline
        \diagbox{\hspace{10pt}}{\hspace{10pt}}& \textbf{Precision} & \textbf{Recall}  & \textbf{F1}\\\hline
        Less & 0.77 & 0.29 & 0.42 \\\hline 
        Equi & 0.78 & 0.96 & 0.86 \\\hline
        More & 0.23 & 0.08 & 0.12 \\\hline
        \end{tabular}
        
        \label{tab:conf_matrix_CC_metric_ori}
    \end{subtable}
\end{table}

In order to comprehensively evaluate this metric, we proceed to compute the precision, recall, and F1-scores for each category.
The confusion matrix along with pertinent metrics are displayed in \autoref{tab:conf_matrix_CC_ori_all}. Although the F1-score of the \textbf{\textit{Equi}} is relatively high (0.86), F1-scores of \textbf{\textit{Less}} and \textbf{\textit{More}} are 0.42 and 0.12 respectively, yielding a macro F1-score of 0.47. 

As shown in \autoref{tab:conf_matrix_CC_ori_all}, 27.4\% (353/1287) of files exhibit a significantly enhanced or diminished level of understandability after decompilation. Nevertheless, these occurrences assume considerable significance, as they imply possible failure in the designing of heuristic algorithms of decompilation for selecting appropriate syntactic representations for the identical semantics. Specifically, Cognitive Complexity only achieves a recall of 0.29 for \textbf{\textit{Less}} understandable decompiled files. Stated differently, Cognitive Complexity exhibited the capability to identify merely 79 instances among the total 270 files of this category, concurrently mistook 181 of the remaining 191 files as \textbf{\textit{Equi}}. 
The utilization of Cognitive Complexity in the evaluation of decompilation understandability, if possible, may introduce the potential for developers to inadvertently overlook these 191 files. This oversight warrants serious consideration, as these files may harbor indications of failure designs contributing to the increased complexities in the decompiled code.
%Once Cognitive Complexity is used in assessing the understandability of decompilation, it is highly possible that these 181 files may be overlooked by developers, though they may contain hints for the failure design that complicate the generated code.

%, which means that the metric does not perform well in predicting these classes. 

% Balanced Accuracy is an average of recalls. 

Despite the low recall of Cognitive Complexity in recognizing \textbf{\textit{Less}} and \textbf{\textit{More}} understandable files, Cognitive Complexity achieves a precision of 0.77 in recognizing \textbf{\textit{Less}} understandable files. This fact shows the potential of Cognitive Complexity as a measure of decompiled code understandability. Our further investigation into these correctly recognized \textbf{\textit{Less}} understandability decompiled files reveals that 81.4\% of these files contains more intricate nested code structures (refer to Pattern P1 in Section~\ref{sec:patterncasestudy}) than the original counterparts.
However, it is also not always the case.
There also exists some nested code structures whose understandability can not be precisely assessed with Cognitive Complexity, as exampled in \autoref{fig:kickCommand}. This figure shows a decompilation by \project{FernFlower} of a code snippet from the \project{Bukkit} project. In this decompilation, the nested $if$ statements \code{condition2} and \code{condition3} are placed within the $if$ statement \code{condition1}, increasing the block depth and thereby complicating understandability. While this discrepancy in understandability is reflected in the Cognitive Complexity scores—5 for the original source and 8 for the decompiled code—the difference remains within our threshold (set to 3). Eventually, this instance results in a misclassification from \textbf{\textit{Less}} to \textbf{\textit{Equi}}. 

These observations underscore the effectiveness of Cognitive Complexity in capturing the complexity of the control flow. However, when it comes to some nested conditional or loop structures, Cognitive Complexity fails to accurately reflect the precise impact of these structures on code understandability in decompilation.
%Nevertheless, the latter exhibits a more intricate code structure, which remain undifferentiated under the current evaluation criteria. It suggests that Cognitive Complexity can still fail in recognizing the de facto understandability of code with deeply nested conditional or loop statements, and there is room for improvement for Cognitive Complexity.

\begin{figure}[htbp]
\centering
    \begin{subfigure}{.22\textwidth}
        \begin{lstlisting}[style=Java, columns=flexible, basicstyle=\scriptsize]
if (condition1) {      //+1
    ... return false;
}
...
if (condition2) {      //+1
    ...
    if (condition3){   //+2
        ...
    }
    ...
} else {               //+1
    ...
}
return true;
// Cognitive Complexity = 5
        \end{lstlisting}
    \vspace{0.1cm}
        \caption{Original}\label{lst:kickCommand-ori}
    \end{subfigure}
    \hspace{0.1cm}
    \begin{subfigure}{.22\textwidth}
        \begin{lstlisting}[style=Java, columns=flexible, basicstyle=\scriptsize]
if (!condition1) {        //+1
    ...
    if (condition2) {     //+2
        ...
        if (condition3) { //+3
            ...
        }
        ...
    } else {              //+1
        ...
    }
    return true;
} else {                  //+1
    ... return false;
}  // Cognitive Complexity = 8
        \end{lstlisting}
    \vspace{0.1cm}
        \caption{\project{Fernflower}}\label{lst:kickCommand-fern}
    \end{subfigure}
\caption{Cognitive Complexity scores for original \texttt{org.\allowbreak{}bukkit.\allowbreak{}command.\allowbreak{}defaults.\allowbreak{}KickCommand} and corresponding code generated by \project{Fernflower}.}
\label{fig:kickCommand}
\end{figure}

\begin{result}

Cognitive Complexity demonstrates relatively acceptable precision while low recall in recognizing decompiled code snippets exhibiting diverse understandability relative to their respective source code. 
%Nested loop or conditional structures may be an important reason for the low recall. // Other factors affecting code understandability that may be an important reason for the low recall. 
\end{result}

\begin{figure*}[htbp]
\includegraphics[width=0.33\textwidth]{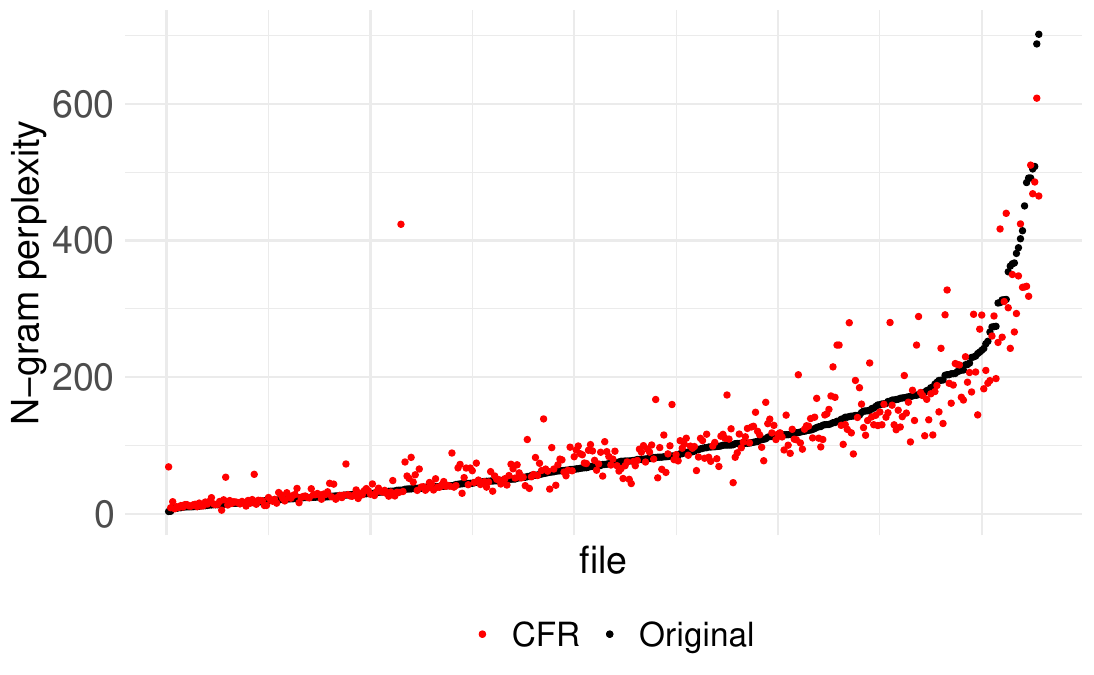}
\includegraphics[width=0.33\textwidth]{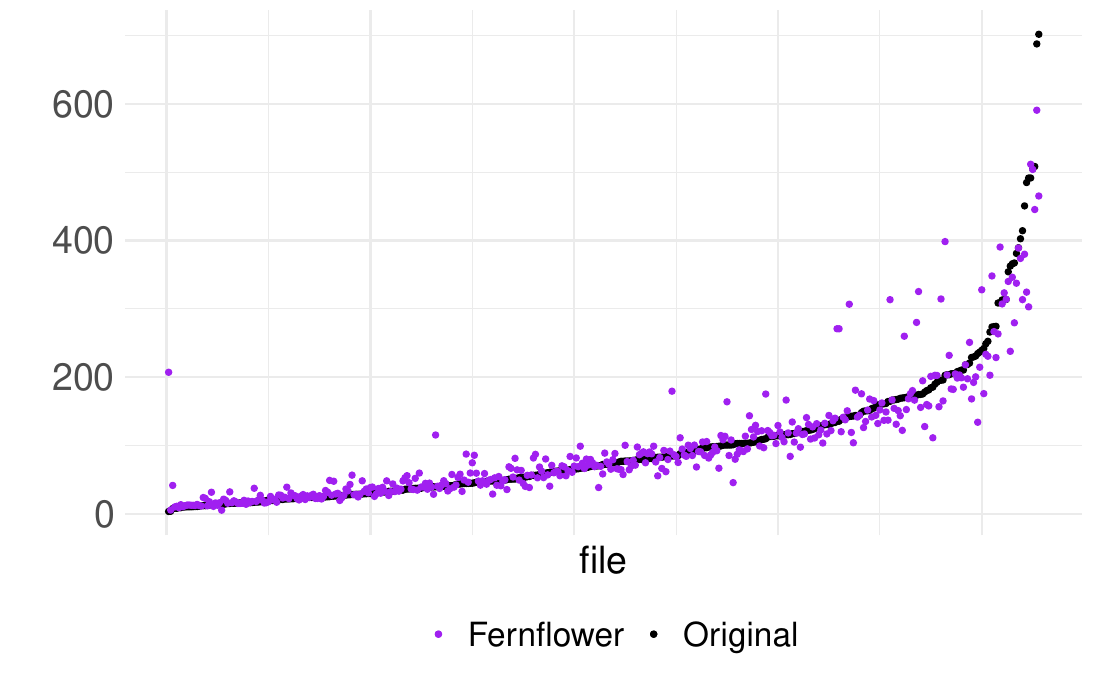}
\includegraphics[width=0.33\textwidth]{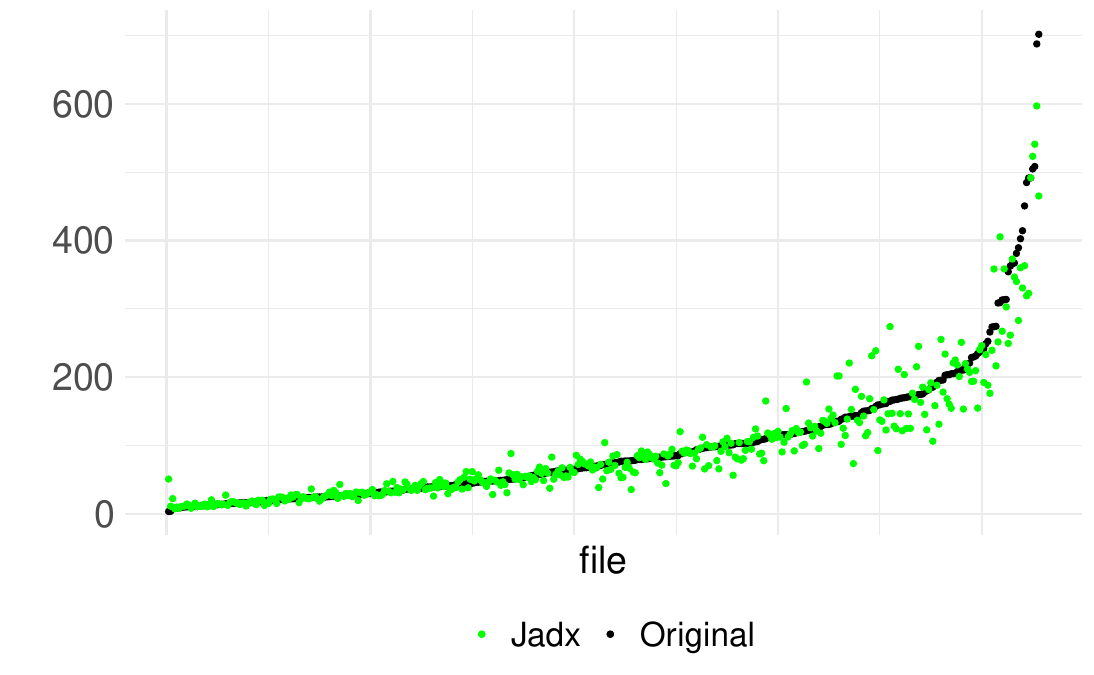}
\caption{Scatter plots of n-gram perplexity scores of original files and corresponding decompiled files} \label{fig:scatter_pp}
\end{figure*}

\subsection{RQ4: how can language-model-based metrics assess the understandability of both decompiled code and their corresponding source code?} \label{subsec:rq4}

To address RQ4, \autoref{fig:scatter_pp} presents perplexity scores of decompiled files and their corresponding original files with an outlier removed, as well. The only outlier is observed in a source file from the project \project{Imaging}, characterized by an original perplexity of 3218.9. The perplexities of its corresponding decompiled code, generated by \project{CFR}, \project{FernFlower}, and \project{Jadx}, manifest as 3134.0, 3153.7, and 4502.6, respectively.
Except this outlier, the perplexity metrics for the remaining source files span a broad spectrum, ranging from 3.7 to 701.5, with an average value of 99.1. With decompiler \project{CFR}, \project{Fernflower}, and \project{Jadx}, the perplexity of respective decompiled files falls within the range of 5.6 to 608.1, 5.6 to 590.5, and 7.6 to 596.7, resulting in average values of 101.0, 100.2, and 94.5.

It's noteworthy that no significant increases or decreases are observed between scores of original files and files generated by different decompilers. To evaluate perplexity, we compute the confusion matrix and relative metrics of the relative understandability indicated by perplexity with reference to the annotated dataset. We adopt a dynamic approach that utilizes a ratio, rather than a constant, to determine the bounds of equivalent understandability. This approach is motivated by the considerable difference in perplexity scores exhibited by files with distinct characteristics. 
Specifically, we employ the following formula:
\[
U(x,ori) = 
\begin{cases}
    Less, & \text{if } x > (1+t)*ori \\
    Equi, & \text{if } (1-t)*ori \leq x \leq (1+t)*ori \\
    More, & \text{if } x < (1-t)*ori
\end{cases}
\]
where $x$ represents the perplexity score of the decompiled code, $ori$ represents the perplexity score of the original. 

We determine a ratio of 0.27 as our selected threshold, which yielding a highest macro F1-score of 0.40.

% 0.0 - 0.35
% The rationale for adopting this ratio is rooted in its effectiveness in achieving meaningful distinctions in understandability assessments, particularly with regard to the unique attributes of perplexity scores across a diverse range of code characteristics.

\begin{table}[thbp]
    \caption{Confusion Matrix and metrics of relative understandability indicated by perplexity.}
    \label{tab:conf_matrix_ngram_ori_all}
    \centering
    \begin{subtable}[htbp]{0.47\linewidth}  \footnotesize
        \centering
        \caption{Confusion Matrix}
        \setlength{\tabcolsep}{0.35em}
            \begin{tabular}{|c|c|c|c|}
            \hline
                \diagbox{\textbf{Predict}}{\textbf{Actual}}  & Less & Equi & More \\\hline
                Less & 68 & 121 & 2 \\
                \hline
                Equi & 184 & 753 & 70 \\
                \hline
                More & 18 & 60 & 11 \\
                \hline
            \end{tabular}
        \label{tab:conf_matrix_ngram_ori}
    \end{subtable}
    \hfill
    \vspace{0.2cm}
    \begin{subtable}[htbp]{0.49\linewidth}  \footnotesize
    \centering
    \caption{Metrics}
    \setlength{\tabcolsep}{0.35em}
        \begin{tabular}{|c|c|c|c|}
        \hline
        \diagbox{\hspace{10pt}}{\hspace{10pt}}& \textbf{Precision} & \textbf{Recall}  & \textbf{F1}\\\hline
        Less & 0.36 & 0.25 & 0.30 \\\hline 
        Equi & 0.75 & 0.81 & 0.78 \\\hline
        More & 0.12 & 0.13 & 0.13 \\\hline
        % Overall & ? & ? & ? \\\hline
        \end{tabular}
    \label{tab:conf_matrix_ngram_metric_ori}
    \end{subtable}
\end{table}

The results are shown in \autoref{tab:conf_matrix_ngram_ori_all}. Perplexity exhibits several phenomenons similar to that of Cognitive Complexity, albeit more serious. The F1-scores for \textbf{\textit{Less}} and \textbf{\textit{More}} understandable decompiled files are 0.30 and 0.13 respectively. Specifically, perplexity only achieves a precision of 0.36 and a recall of 0.25 for \textbf{\textit{Less}} understandable decompiled files, and the macro F1-score is only 0.40, which suggests that perplexity may not be well-aligned with the code understandability of decompilation.
% might not be appropriate for this scenario. 
% implies that the metric may not be well-aligned with the intricacies of the scenario

%While our analysis of files that are predicted as \textbf{\textit{Less}} understandable by perplexity reveals some noteworthy findings. Among these files, P3(excessively long lines) is found in 27 of them, and P2(omitted parentheses in expressions with mixed operators) is observed in 19 of them. 
% which are not observed in their corresponding original files. 
These figures reveal that perplexity is difficult to capture the unnaturalness of the code, suggesting that it may require further refinement or be more effective in identifying certain types of unnatural code patterns. This finding underscores the complexity of assessing code understandability and highlights the challenges in automating this process effectively.

\begin{result}
Perplexity demonstrates both low precision and recall in recognizing decompiled code snippets exhibiting diverse understandability relative to their respective source code. %Unnatural code structures, such as inlined assignments in expressions, may explain the low precision and recall. // multiple factors that impact perplexity may explain the low precision and recall. 
\end{result}

% \subsection{RQ3: how is the understandability of decompiled code from the perspective of decompilers?}\label{subsec:rq3}

% To answer this question, we first analyze the performance of two metrics from the perspective of decompilers. Subsequently, we conduct a manual analysis on specific decompilers and summarize the code that impact relative understandability.

% \find{Conclusion}{The understandability of code generated by the same compiler tend to share similar evaluations with Cognitive Complexity and Perplexity, whereas the only exception lies in cases such as omitted optional braces after conditional or
% loop statements, and inlined assignments in expressions.}

% \subsection{RQ4: how is the decompiled code understandability from the perspective of projects?}

% \find{Conclusion}{The understandability of code in the same project decompiled by all the three decompilers tend to share similar evaluations with Cognitive Complexity and Perplexity, whereas the only exception lies in cases such as direct usage of numerical literals in methods.}

\section{Case Study}
\label{sec:casestudy}
Subsequently, we conduct an in-depth analysis of the causes contributing to the understandability disparities between the decompiled code and original source code. We further categorize and present a comprehensive summary of these causes based on their structural manifestations, specifically in terms of code patterns.
\subsection{Code Patterns Leading to Understandability Disparities}
\label{sec:patterncasestudy}
These patterns, named as P1-P6, are summarized in \autoref{tab:patterns}. We offer six real cases and corresponding analysis to illustrate the impact of these patterns. These case studies shed light on the influence of these patterns on relative understandability.

% Our manual analysis on decompiled code and corresponding original source code has revealed several types of code patterns that are semantically equivalent, while vary in syntax and understandability. 
% In summary, decompiled code understandability is primarily influenced by the following four aspects:
% \begin{enumerate*}
% \item \textbf{Code Structure (P1)}: The structure of the code, including complex control flow patterns, especially deeply nested conditional statements, plays a substantial role in influencing understandability.
% \item \textbf{Expression (P2 and P5)}: Expressions within the code, including complex expressions without parentheses and nested assignment expressions, can affect understandability. % P2 can obscure the intent of expressions, especially when developers are unfamiliar with the operator precedence, thereby impeding code understandability. 
% \item \textbf{Formatting (P3 and P4)}: The formatting of the code, such as excessively long lines and omitted optional braces, can decrease understandability. 
% \item \textbf{Constant (P6)}: Code that includes a number of numerical literals, especially when they are not well-documented, reduces understandability.
% \end{enumerate*}

\begin{table}[htbp]
\caption{Code patterns that impact code understandability.}
\footnotesize
    \centering
    \begin{tabular}{|c|l|}
	\hline
	\textsc{\textbf{\#P}} & \textsc{\textbf{Description}} \\ \hline
        P1 & Deeply nested conditional or loop statements. \\ \hline
        P2 & Omitted parentheses in expressions with mixed operators. \\ \hline
        P3 & Excessively long lines. \\ \hline
        P4 & Omitted braces after conditional or loop statements. \\ \hline
        P5 & Inlined assignments in expressions. \\ \hline
        P6 & Numerical literals in methods directly. \\ \hline
    \end{tabular}
    \label{tab:patterns}
\end{table}

% previous RQ3 from the perspective of decompilers   RQ4 from the perspective of projects
% we analyze code patterns identified in decompiled files of distinct projects. 
% To study how Cognitive Complexity performs with respect to various decompilers, we display confusion matrix by decompiler in \autoref{tab:conf_matrix_CC_ori_per_decompiler}. 

% While across all three decompilers, the recall of \textbf{\textit{Less}} and \textbf{\textit{More}} is notably low, which indicates that Cognitive Complexity is less reliable when it comes to distinguishing files that are either more or less understandable.

\begin{figure}[htbp]
\centering
    \begin{subfigure}{.22\textwidth}
        \begin{lstlisting}[style=Java, basicstyle=\scriptsize]
if (args.length < 2) {
    ... return true;
} 
if (player == null) {
    ... return true;
}
if (condition3) {
    [...] return true;
}

if (condition4) {
    ...
}
[...]
        \end{lstlisting}
    \vspace{0.1cm}
        \caption{Original}\label{lst:effectCommand-ori}
    \end{subfigure}
    \hspace{0.1cm}
    \begin{subfigure}{.22\textwidth}
        \begin{lstlisting}[style=Java, basicstyle=\scriptsize]
if (args.length < 2) {
    ... return true;
} @@else@@ {
    if (player == null) {
        ... return true;
    } @@else@@ if (!condition3) {
        @@if (condition4) {@@
            ...
        @@}@@
        [...]
    } @@else@@ {
        [...] return true;
    }
}
        \end{lstlisting}
    \vspace{0.1cm}
        \caption{\project{Fernflower}}\label{lst:effectCommand-fern}
    \end{subfigure}
\caption{Excerpts of \texttt{org.\allowbreak{}bukkit.\allowbreak{}command.\allowbreak{}defaults.\allowbreak{}EffectCommand}, original and decompiled by \project{Fernflower}.}
\label{fig:effectCommand}
\end{figure}

\noindent\textbf{P1: deeply nested conditional or loop statements.}
The term ``deeply" in P1 means that these nested structures should have a minimum depth of 3. 
To illustrate P1, \autoref{fig:effectCommand} provides a comparison between a subset of conditional statements within the original source file from project \project{Bukkit} and the decompiled file generated by \project{Fernflower}.
The source file in \autoref{lst:effectCommand-ori} depicts a series of $if$ blocks where developers conclude each block with $return$ statements. This practice is well-known for reducing branch depth in Java and other programming languages for better code comprehension.  However, as shown in \autoref{lst:effectCommand-fern}, \project{Fernflower} reorganized these 1-depth conditional statements as a huge and deeply nested $if$ statement, which significantly amplified the difficulty in code comprehension. Retaining the original $if$ statements, as seen in \autoref{lst:effectCommand-ori}, from the  presents a non-trivial challenge for Java decompilers. The main reason is that $if$ statements with $return$ and nested $if$ statements with $else$ share the identical bytecode. Java decompilers face significant challenges in determining the optimal generation strategy for these structures. Any oversight in this process can adversely affect the understandability of the decompiled code. %Additionally, the complexity and depth of the conditional statement generated by \project{Fernflower} would increase if the source code contained a greater number of 1-depth conditional statements.

In addition, P1 can be frequently observed in quite a number of decompiled files. Particularly, among 300 decompiled files from project \project{Bukkit}, over one-fifth of these files are confirmed to exhibit characteristics associated with P1.

\noindent\textbf{P2: omitted parentheses in expressions with mixed operators.}
To illustrate P2, we provide a snippet of the original source code from \project{Imaging}, alongside the code decompiled by \project{CFR} in \autoref{fig:mybit}. In the Java programming language, various operators, including arithmetic operators such as `-' and binary operations such as `\&' and `$>>$' used in \autoref{fig:mybit}, have their own execution priorities. Notably, decompilers may opt to eliminate parentheses in order to simplify the generated code. However, this practice is not without its drawbacks, as parentheses conventionally serve the purpose of aiding readers in deducing the proper execution order of a given expression. The absence of such parentheses can potentially lead to confusion for readers. This dilemma necessities decompilers to carefully design their strategies, determining whether to include parentheses for comprehensibility, or just to omit them for simplicity under various scenarios. %It becomes evident that the decompiled code is less understandable than its original counterpart. The issue that contributes to this decrease in code understandability is the ignorance of parentheses of expressions containing various binary operators. This practice can obfuscate the meaning of these expressions, making it challenging for developers to discern their intent and functionality. It is important to note that an exception is found in cases involving sequential common arithmetic operators, which adhere to the same precedence rules as they do in mathematics.

\begin{figure}[htbp]
\centering
    \begin{subfigure}{.46\textwidth}
        \begin{lstlisting}[style=Java, basicstyle=\scriptsize]
sample = sampleMask & (bitCache >> (bitsInCache - sampleBits));
        \end{lstlisting}
    \vspace{0.1cm}
        \caption{Original}\label{lst:mybit-ori}
    \end{subfigure}
    \begin{subfigure}{.46\textwidth}
        \begin{lstlisting}[style=Java, basicstyle=\scriptsize]
sample = sampleMask @@&@@ this.bitCache @@>>@@ this.bitsInCache @@-@@ sampleBits;
        \end{lstlisting}
    \vspace{0.1cm}
        \caption{\project{CFR}}\label{lst:mybit-cfr}
    \end{subfigure}
\caption{Excerpts of \texttt{org.\allowbreak{}apache.\allowbreak{}commons.\allowbreak{}imaging.\allowbreak{}common.\allowbreak{}mylzw.\allowbreak{}MyBitInputStream}, original and decompiled by \project{CFR}.}
\label{fig:mybit}
\end{figure}

P2 is frequently observed in the decompilation of projects characterized by low-level numerical and bit calculations, such as projects \project{Imaging} and \project{Joda Time}. Particularly, 47 out of 258 decompiled files in project \project{Imaging} present to contain P2.

\begin{figure}[htbp]
\centering
    \begin{subfigure}{.46\textwidth}
        \begin{lstlisting}[style=Java, basicstyle=\scriptsize]
if (end < start) {
    size = maxElements - start + end;
} else if (end == start) {
    size = full ? maxElements : 0;
} else {
    size = end - start;
}
        \end{lstlisting}
    \vspace{0.1cm}
        \caption{Original}\label{lst:fifo-ori}
    \end{subfigure}
    \begin{subfigure}{.46\textwidth}
        \begin{lstlisting}[style=Java, basicstyle=\scriptsize]
size = this.end < this.start ? this.maxElements - this.start + this.end : (this.end == this.start ? (this.full ? this.maxElements : 0) : this.end - this.start);
        \end{lstlisting}
    \vspace{0.1cm}
        \caption{\project{CFR}}\label{fifo:mybit-cfr}
    \end{subfigure}
\caption{Excerpts of 
\texttt{org.\allowbreak{}apache.\allowbreak{}commons.\allowbreak{}collections4.\allowbreak{}queue.\allowbreak{}CircularFifoQueue}, original and decompiled by \project{CFR}.}
\label{fig:fifo}
\end{figure}

\noindent\textbf{P3: excessively long statements.} 
To illustrate P3, we provide a snippet from project \project{Collections} in \autoref{fig:fifo}. In the original code snippet, a straightforward set of conditional statements is utilized to determine the value of the 'size' variable based on various conditions related to the 'start' and 'end' variables. While \project{CFR} demonstrates the tendency to compact expressions, leveraging nested ternary operators to represent the same logic. As the complexity of the conditions increases, the resulting nested ternary expressions could become increasingly complex, which leads to excessively long statements and do harm to code understandability.

The prevalence of P3 is principally rooted in the common-used optimization strategies for generating concise conditional operator expressions when confronted with multiple branches assigning values to the same variable. This preference for brevity aims to make the decompiled code easier to grasp.

However, when there are multiple branches in the code or the conditions become intricate, this simplification can lead to excessively long and convoluted statements. As a result, the understandability of the decompiled code suffers. This highlights a delicate balance between keeping the code concise and ensuring it remains clear and understandable.

In simpler terms, decompilers often try to make code shorter and easier to understand by condensing complex conditions. But sometimes, this can make the code too long and hard to follow, showing the importance of finding the right balance between simplicity and clarity in decompiled code.

\begin{figure}[htbp]
\centering
    \begin{subfigure}{.46\textwidth}
        \begin{lstlisting}[style=Java, basicstyle=\scriptsize]
if (Character.toUpperCase(c1) != Character.toUpperCase(c2)
     && Character.toLowerCase(c1) != Character.toLowerCase(c2)) {
    return false;
}
        \end{lstlisting}
    \vspace{0.1cm}
        \caption{Original}
    \end{subfigure}
    \begin{subfigure}{.46\textwidth}
        \begin{lstlisting}[style=Java, basicstyle=\scriptsize]
if @@(Character.toUpperCase(c1) == Character.toUpperCase(c2) || Character.toLowerCase(c1) == Character.toLowerCase(c2)) continue@@;
return false;
        \end{lstlisting}
    \vspace{0.1cm}
        \caption{\project{CFR}}\label{lst:charseq-cfr}
    \end{subfigure}
\caption{Excerpts of \texttt{org.\allowbreak{}apache.\allowbreak{}commons.\allowbreak{}codec.\allowbreak{}binary.\allowbreak{}CharSequenceUtils}, original and decompiled by \project{CFR}.}
\label{fig:charseq}
\end{figure}

% even label break   simplifying the control flow?
% the choice and decision between simplicity and clarity
\noindent\textbf{P4: omitted braces after conditional or loop statements.}
To exemplify P4, we provide a comparative analysis between the original source code and the code generated by \project{CFR} within the context of a $while$ loop located in a method of the class belonging to project \project{Codec}, as presented in \autoref{fig:charseq}. Notably, decompiler \project{CFR} alters the loop condition by negating it and utilizes the $continue$ statement with omitted braces following the $if$ condition, as shown in \autoref{lst:charseq-cfr}. 
Decompilers are designed with the goal of producing code that retains semantic equivalence to the original source code while ensuring simplicity and clarity. Omitted optional braces within the code can be perceived as a strategy of decompilers to reduce verbosity and enhance the compactness of the decompiled code. However, this practice has the side effect of obscuring the boundaries of code blocks, potentially diminishing the clarity in discerning where a block begins and ends. This, in turn, can have impact on code understandability, especially when the conditional logic is intricate.
% it is essential to recognize that 
% and harder to understand the intended logic.

% The omission of braces and the concatenation of the $continue$ statement with the $if$ condition on a single line can lead to longer lines of code, which, in turn, can hinder the ease of code comprehension. This behavior is particularly noteworthy and is often observed in \project{CFR}'s decompilation output.

% P4 is more frequently observed in files generated by \project{CFR}. 
% P4 is also observed in original source code
% Upon closer examination of  and comparison with the original files, 
% It's noteworthy that P4 is also observed in original source code, and P5 is also frequently found in code generated by \project{Jadx}.

\begin{figure}[tbp]
\centering
    \begin{subfigure}{.4\textwidth}
        \begin{lstlisting}[style=Java, basicstyle=\scriptsize]
char mappedChar = map(str.charAt(index));
if (index > 1 && mappedChar != '0') {
    final char hwChar = str.charAt(index - 1);
    if ('H' == hwChar || 'W' == hwChar) {
        [...]
    }
}
        \end{lstlisting}
    \vspace{0.1cm}
        \caption{Original}\label{lst:soundex-ori}
    \end{subfigure}
    \begin{subfigure}{.4\textwidth}
        \begin{lstlisting}[style=Java, basicstyle=\scriptsize]
char hwChar;
char mappedChar = map(str.charAt(index));
if (index > 1 && mappedChar != '0' && @@('H' == (hwChar = str.charAt(index - 1)) || 'W' == hwChar))@@ {
    [...]
}
        \end{lstlisting}
    \vspace{0.1cm}
        \caption{\project{Jadx}}\label{lst:soundex-jadx}
    \end{subfigure}
\caption{Excerpts of \texttt{org.\allowbreak{}apache.\allowbreak{}commons.\allowbreak{}codec.\allowbreak{}language.\allowbreak{}Soundex}, original and decompiled by \project{Jadx}.}
\label{fig:soundex}
\end{figure}

\noindent\textbf{P5: inlined assignments in expressions.}
Concerning P5, \autoref{lst:soundex-jadx} represents another typical and prevalent optimization strategy provided by Java decompilers, as illustrated in line 3. This optimization strategy inlines trivial assignments in associated expressions, thereby positioning assigned variables in proximity to their usage within the expression, consequently enhancing comprehensibility and simplicity. However, as depicted in \autoref{lst:soundex-jadx}, the inappropriate application of this optimization may conversely compromise code comprehensibility. In this case, variable `hwChar' is instantiated through a rather complex manner using `charAt' invocation. Subsequently, although `H' and `W' are both compared to `hwChar', they exhibit disparate formats. With these two points, this inlining practice provided by \project{Jadx} significantly hindered the code understandability and confused the code readers. In addition to considerations related to assignment complexity and expression similarity, numerous other factors necessitate attention in the designing of this optimization strategy. Java decompiler developers should exercise clear awareness of these considerations. Regrettably, such awareness is often lacking in practice.
%This compound statement inlines an assignment within a conditional expression, while this statement can potentially be misinterpreted as the equality operator. This misinterpretation can introduce challenges in terms of code readability and significantly diminish the overall code's understandability.

%Furthermore, the use of such intricate compound statements can pose difficulties when developers attempt to identify the precise point of assignment within these complex expressions, especially when the variable is subsequently referenced in other statements. This complexity can impede the code's understandability and hinder the ease of tracking assignment points within the code.

\begin{figure}[htbp]
\centering
    \begin{subfigure}{.43\textwidth}
        \begin{lstlisting}[style=Java, basicstyle=\scriptsize]
private static final int gridSize = 1 << 24;
public BlockIterator(...) {
  [...]
  secondError = floor(secondd * gridSize);
  [...]}
        \end{lstlisting}
    \vspace{0.1cm}
        \caption{Original}\label{lst:blockIterator-ori}
    \end{subfigure}
    \begin{subfigure}{.43\textwidth}
        \begin{lstlisting}[style=Java, basicstyle=\scriptsize]
private static final int gridSize = @@16777216@@;
public BlockIterator(...) {
  [...]
  this.secondError = NumberConversions.floor(secondd * @@1.6777216E7@@);
  [...]}
        \end{lstlisting}
    \vspace{0.1cm}
        \caption{\project{Fernflower}}\label{lst:blockIterator-fern}
    \end{subfigure}
\caption{Excerpts of \texttt{org.\allowbreak{}bukkit.\allowbreak{}util.\allowbreak{}BlockIterator}, original and decompiled by \project{Fernflower}.}
\label{fig:blockIterator}
\end{figure}

% 20231218  It seems that the authors confuse optimizations done by the compiler with computations done by the decompiler. The actual problem is that the decompiler is not able to undo certain code optimizations done by the compiler including constant propagation.
\noindent\textbf{P6: numerical literals instead of constants.}
Within source code, developers typically create a multitude of well-named constants that are used across their projects. During the compilation process to bytecode, these constants, particularly the numerical constants, are invariably substituted with concrete numerical literals by compilers with constant propagation optimization. Existing decompilers frequently encounter challenges in retaining these numerical literals to the corresponding constants as source code. As shown in \autoref{fig:blockIterator}, \project{FernFlower} used $1.67777216E7$ instead of $gridSize$ in line 4 of \autoref{lst:blockIterator-fern}. This deviation impedes the clarity and understandability of the decompiled code. 
Furthermore, the situation can be more complex for expressions comprised of the calculation using multiple numerical constants. Java compilers often employ constant folding optimization, which simplifies the code by computing the results of these expressions at compile-time. 
% rather than retaining the procedural formulae. 
As shown in line 1 of \autoref{lst:blockIterator-fern}, \project{FernFlower} used $167777216$, i.e., the result of $1<<24$ rather than recovering the original expression. This renders the decompiled code, particularly with regard to numerical constants, significantly more difficult to comprehend. % This transformation further obfuscates the logic and intent behind the code, making the decompiled output significantly more difficult to comprehend, particularly when numerous such optimizations are involved.

Addressing these understandability issues presents a formidable challenge. Decompilers must decide whether to retain each numerical literal in association with its potential corresponding constant. This requires sophisticated analysis to identify constants and expressions that were optimized by the compiler and map them back to their original named constants and expressions.
% Addressing these understandability issues in decompiled code presents a formidable challenge. To mitigate the potential understandability impediments in P6, it is imperative for Java decompilers to meticulously determine not only whether to retain each numerical literal in association with its potential corresponding constant 
To mitigate the potential understandability impediments of P6, it is imperative for Java decompilers to implement advanced heuristics and analysis techniques. These should aim to recognize common constants and patterns in expressions, reverse engineer the optimizations performed by the compiler, and replace literals with meaningful constant names wherever possible.

%to determine not only whether to retain each numerical literal to the potential corresponding constant variable, but also whether the constant propagation optimization should be applied on a specific calculation involving multiple numeric constants.

%Numerical literals lack the self-documentation that well-named constants or variables provide, and are often used without any accompanying explanation or context. Things may become even worse when some variables are designed from the combination of several constants. Decompilers are tended to use constant propagation to get the eventual results without providing any processes . 
%This makes it difficult for developers to understand what these numbers represent or why they are used.
%Whether a contant should be translated to an predefined contant variable

Presently, P6 has been observed across many projects, including \project{Codec}, \project{Imaging}, \project{Lang} and \project{Joda Time}. It is evident that P6 has the potential to occur in any other projects that frequently define and use numerical constants. 

% It is noteworthy that in cases where multiple lines are manually separated from a long line in the original source files, all decompilers share the same behavior of merging them into one. This observation suggests that decompilers have not noticed their strategies may decrease the code understandability from the perspective of formatting before.

% In summary, our analysis reveals project-specific variations in the prevalence of code patterns within decompiled code. These variations are influenced by the individual characteristics of projects and specific coding styles employed in projects.

% In summary, code may suffer from a lack of clarity due to the absence of parentheses in expressions involving different binary operators. This issue has the potential to impede the understanding of the code.

% although it is worth noting that arithmetic operators that follow standard mathematical precedence rules are not as affected by this concern.

%In order to investigate the universality and diversity of these patterns, we proceeded to perform a comprehensive statistical analysis on the files with these patterns generated by each decompiler. The result is presented in \autoref{tab:patterns-per-dec}. Patterns including P1, P2, P5, and P6 can be observed in the output of all three decompilers. While P1, P2, P3, P4 along with P6, are either not observed or observed to a limited extent in specific decompilers. This finding emphasizes the potential for decompilers to mitigate or entirely eliminate these particular patterns.

\subsection{Prevalence\&Severity of {Code Patterns}}
% Based on the outcomes consolidated from our RQ2, we obtain several semantically equivalent code patterns in decompilation, i.e., P1-P6 that exhibit significant syntactic disparities impacting code understandability. Furthermore, the outcomes gleaned from RQ3 and RQ4 reveal that these identified patterns pose a considerable likelihood of introducing ambiguity into the code understandability in terms of Cognitive Complexity and perplexity.
Although these patterns are summarized based on our insights, it is also essential to conduct a quantitative analysis on the prevalence and severity of these patterns. This analysis is aimed at giving a more comprehensive understanding of these code patterns.
To this end, \autoref{tab:patterns-counts} list the distribution of the total six patterns in \autoref{tab:patterns} over all the 1287 decompiled files contained across the three decompilers and the twelve projects, respectively. 
% Each cell in \autoref{tab:patterns-counts} represented as \emph{a/b/c} represents there are \emph{a} decompiled files that contain such code pattern, among which \emph{b} files exhibit lower understandability relative to their source code files. Within these \emph{b} files, there are further \emph{c} files whose \textbf{\textit{Less}} understandability relative to the source files can be neither recognized by Cognitive Complexity nor perplexity. 
Here we include only the decompiled files with \textbf{\textit{Less}} understandability because all the six patterns tended to diminish the understandability of the code.
% To note, the last column (\textsc{\textbf{\#(P1\ldots P6)}}) in the tables stand for the sum of files that contain any one of the six patterns across the three decompilers and the twelve projects.
% 20231218 SANER review2 The caption should explain what those slash-separated numbers are or, even better, divide the columns into sub-columns with explicit captions for each of them.  

% These code patterns are now enumerated in \autoref{tab:patterns}.
\begin{table}[htbp]\scriptsize
\caption{Number of decompiled files with patterns listed in \autoref{tab:patterns} (Each cell denoted as \emph{a/b/c} represents that there are \emph{a} decompiled files that contain such code pattern, among which \emph{b} files exhibit lower understandability relative to their source code files. Within these \emph{b} files, there are further \emph{c} files whose \textbf{\textit{Less}} understandability relative to the source files can be neither recognized by Cognitive Complexity nor perplexity. \textsc{\textbf{\#(P1\ldots P6)}} stands for the sum of files that contain any one of the six patterns.). }
    \label{tab:patterns-counts}
    \setlength{\tabcolsep}{0.35em}
% \scriptsize
    \centering
    
\begin{subtable}[htbp]{\linewidth}
        \centering
        \caption{By decompilers}
    \begin{tabular}{|l|c|c|c|c|c|c|c|c|}
	\hline
	\textsc{\textbf{Dec}} & \textsc{\textbf{\#P1}} & \textsc{\textbf{\#P2}} & \textsc{\textbf{\#P3}} & \textsc{\textbf{\#P4}} & \textsc{\textbf{\#P5}} & \textsc{\textbf{\#P6}} & \textsc{\textbf{\#(P1\ldots P6)}}\\ \hline
        \project{CFR}        & 5/4/0    & 47/43/30 & 23/17/11 & 46/36/20 & 23/21/16 & 13/12/11 & 125/105/71 \\ \hline
        \project{Fern}       & 73/56/4  & 40/36/22 & 31/22/9  & 0/0/0    & 0/0/0    & 13/12/5  & 135/105/37 \\ \hline
        \project{Jadx}	     & 42/24/6  & 8/5/4    & 28/24/15 & 0/0/0    & 4/2/1    & 7/6/5    & 83/57/31   \\ \hline
    \textsc{\textbf{Total}}  & 120/84/10 & 95/84/56 & 82/63/35 & 46/36/20 & 27/23/17 & 33/30/21 & 343/267/139 \\ \hline
    \end{tabular} 
    \label{tab:patterns-per-dec}
    \medskip
\end{subtable}
 % \smallskip,\medskip,\bigskip
\begin{subtable}[htbp]{\linewidth}
        \centering
        \caption{By projects}
    \setlength{\tabcolsep}{0.07em}
    \begin{tabular}{|l|c|c|c|c|c|c|c|c|}
		\hline
		\textsc{\textbf{Project}} & \textsc{\textbf{\#File}} & \textsc{\textbf{\#P1}} & \textsc{\textbf{\#P2}} & \textsc{\textbf{\#P3}} & \textsc{\textbf{\#P4}} & \textsc{\textbf{\#P5}} & \textsc{\textbf{\#P6}} & \textsc{\textbf{\#(P1\ldots P6)}}\\
	\hline
         Bukkit            & $300$ & 60/30/5   & 14/13/10   & 19/14/9   & 12/7/2   & 7/5/5 & 2/2/1  & 99/58/28 \\ \hline
        Codec              & $78$  & 17/16/1   & 6/6/2      & 8/7/0     & 6/6/3    & 7/5/2 & 8/8/4  & 37/34/10 \\ \hline
        Collections        & $96$  & 0/0/0     & 3/3/2      & 0/0/0     & 2/2/0    & 0/0/0 & 0/0/0  & 5/5/2    \\ \hline
        Imaging	           & $258$ & 16/15/2   & 47/47/30   & 12/12/6   & 11/11/5  & 3/3/2 & 12/9/9 & 85/80/45 \\ \hline
        Lang               & $84$  & 8/7/2     & 5/5/3      & 8/8/3     & 5/5/4    & 2/2/1 & 6/6/4  & 31/30/16 \\ \hline
        Joda time          & $213$ & 9/9/0     & 18/9/8     & 18/16/13  & 3/3/3    & 6/6/5 & 5/5/3  & 53/42/30 \\ \hline
        Jsoup              & $33$  & 3/0/0     & 1/0/0      & 3/0/0     & 0/0/0    & 0/0/0 & 0/0/0  & 7/0/0    \\ \hline
        Junit4             & $108$ & 4/4/0     & 0/0/0      & 7/6/4     & 5/4/2    & 1/1/1 & 0/0/0  & 12/11/5  \\ \hline
        Mimecraft          & $6$   & 1/1/0     & 0/0/0      & 0/0/0     & 0/0/0    & 0/0/0 & 0/0/0  & 2/1/0    \\ \hline
        Scribe Java        & $75$  & 0/0/0     & 1/1/1      & 7/0/0     & 1/1/1    & 1/1/1 & 0/0/0  & 10/3/3   \\ \hline
        Spark              & $27$  & 0/0/0     & 0/0/0      & 0/0/0     & 1/1/0    & 0/0/0 & 0/0/0  & 1/1/0    \\ \hline
        DcTest             & $9$   & 2/2/0     & 0/0/0      & 0/0/0     & 0/0/0    & 0/0/0 & 0/0/0  & 2/2/0    \\ \hline
  \textsc{\textbf{Total}}  &$1287$ & 120/84/10 & 95/84/56   & 82/63/35  & 46/36/20 &27/23/17& 33/30/21 & 343/267/139 \\
        \hline
    \end{tabular} 
    \label{tab:patterns-per-project}
\end{subtable}
\end{table}

In general, an examination of 343 decompiled files, constituting 26\% of the total 1287 files, reveals the presence of at least one of the six patterns. Within this subset, each pattern is found in a range of 27 to 120 decompiled files. While the overall incidence rate may not be deemed substantial, a noteworthy observation emerges from the analysis of these 343 files: 267 of them, comprising over three-quarters of the subset, demonstrate \textbf{\textit{Less}} understandability in comparison to their corresponding source code. % or \textbf{\textit{More}}

It is noteworthy that, within the 1287 decompiled files, a total of 270 files exhibit \textbf{\textit{Less}} of understandability. 
This implies that the 343 files associated with the identified patterns constitute 98.9\% (267/270) of instances exhibiting \textbf{\textit{Less}} understandability, leaving the remaining 1.1\% (3/270) attributed to the 951 files devoid of these patterns.

In more detail, P4 is the only pattern observed in the decompiled files generated by merely \project{CFR}, whereas four of the remaining five patterns are observed in files generated by all three decompilers.  Project-wise, all six patterns are observed within the decompiled files of five projects, namely \project{Bukkit}, \project{Codec}, \project{Imaging}, \project{Lang}, and \project{Joda time}. For the observation that the remaining seven projects do not contain all the six patterns, the limited involvement of these projects' files in our experiment may account. For instance, projects \project{Mimecraft}, \project{DcTest}, and \project{Spark}, each characterized by the presence of only two of the six patterns, contain at most 27 decompiled files.
%Among most projects with more than 70 files, including \project{Bukkit}, \project{Codec}, \project{Imaging}, \project{Lang}, and \project{Joda time}, all the code patterns listed in \autoref{tab:patterns} can be observed. This observation reflects the generalization of these patterns, as they are consistently present in decompiled code across these projects. While \project{Collections}, \project{Junit4} and \project{Scribe Java} are exceptions because original files of these projects are relatively simple and lack of specific patterns. 

%severity
In terms of metrics including Cognitive Complexity and perplexity, it is observed that the two metrics inadequately identified the \textbf{\textit{Less}} understandability of 139 decompiled files relative to their corresponding source files, or specifically, 52.0\% (139/267) of decompiled files containing any of the six patterns. Moreover, these 139 decompiled files constitute 72.7\% (139/(181+10)) and 68.8\% (139/(184+18)) of instances where Cognitive Complexity and perplexity fail to identify \textbf{\textit{Less}} understandability relative to the source files, respectively. The impact of these identified patterns on the precision and recall of the two metrics is presumed to be more profound, considering the existence of quite a number of decompiled files not shown in the provided tables, which the two metrics gave different assessment in their relative understandability.
For instance, although only 10 \textbf{\textit{Less}} understandable files with pattern P1 are recognized as \textbf{\textit{Equi}} understandable with both Cognitive Complexity and perplexity, there exist another 61 \textbf{\textit{Less}} understandable files which are recognized as \textbf{\textit{Less}} understandable with Cognitive Complexity while recognized as \textbf{\textit{Equi}} understandable with perplexity. This example also aligns with our previous observation in \autoref{subsec:rq3}, that Cognitive Complexity is able to capture the complexity of control flows to some extent, but may still encounter difficulties in code snippets involving nested conditional or loop statements.

Examining each pattern individually, it is noted that patterns P2-P6 have 55.6\%-73.9\% decompiled files exhibiting \textbf{\textit{Less}} understandability that were neither identified by Cognitive Complexity nor by perplexity. Noteworthy is the exception found in pattern P1, where only 10 out of 84 decompiled files evaded recognition by both metrics. The remaining 74 decompiled files were mainly identified by Cognitive Complexity. This fact aligns with our earlier observations in addressing RQ3, highlighting Cognitive Complexity's relative effectiveness in capturing control-flow complexity.

% Note that several excerpts with P3 can be also observed in the original source code in most projects, in this situation, P3 is not or less observed in their corresponding decompiled code. While optional braces in some original files may be omitted in decompiled code generated, which decrease the relative understandability.

\subsection{Responses from Java decompiler developers}
\label{subsec:developerresponses}
\begin{table}[!t]
    \scriptsize
	\caption{Eleven understandability issues reported to developers.}
	\centering
    \begin{threeparttable}
	\begin{tabular}{|c|c|c|c|c|c|}
	\hline
    \textbf{ID} & \textbf{Decompiler} & \textbf{Pattern}   &\textbf{Newly Reported}  & \textbf{Issue No.}\tnote{1} & \textbf{State}\\\hline
    1  & Jadx  & P1 & $\times$ & 1455 & Fixed\\\hline
    2  & Jadx  & P1 & $\times$ & 1689 & Confirmed\\\hline
    3  & Jadx  & P1 & $\checkmark$ & 2052 & Fixed\\\hline
    4  & FernFlower & P1 & $\checkmark$ & 342096 & Confirmed\\\hline
    5  & FernFlower & P2 & $\checkmark$ & 343614 & Confirmed\\\hline
    6  & CFR   & P2 & $\checkmark$ & 352 & Reported\\\hline
    7  & Jadx  & P3 & $\checkmark$ & 2123 & Confirmed\\\hline
    8  & CFR   & P4 & $\checkmark$ & 354 & Reported\\\hline
    9  & Jadx  & P5 & $\checkmark$ & 2076 & Confirmed\\\hline
    10  & CFR   & P6 & $\checkmark$ & 353 & Confirmed\\\hline
    11  & FernFlower & P6 & $\checkmark$ & 344050 & Confirmed\\\hline

	\end{tabular}
     \begin{tablenotes}
      \item[1] The issues of \project{CFR}, \project{FernFlower}, and \project{Jadx} can be found at \url{https://github.com/leibnitz27/cfr/issues/XXX}, \url{https://youtrack.jetbrains.com/issue/IDEA-XXX}, and \url{https://github.com/skylot/jadx/issues/XXX}, where "XXX" can be replaced with the concrete numbers in \textbf{Issue No.}.
    \end{tablenotes}
     \end{threeparttable}
	\label{tab:bugrevealed}
    \end{table}
% Although P1 received the lowest score among all the evaluated patterns, 
We have reported all the understandability issues to the decompiler developers, which are detailed in Table~\ref{tab:bugrevealed}. Notably, 7 out of the 9 understandability issues have been confirmed or addressed by the developers, with these 7 issues covering 5 of the 6 patterns summarized from the first experiment. These observations suggest that the developers are indeed concerned with the understandability of the decompiled code and acknowledge most of the patterns we identified. 

It is also important to note that in \project{Jadx}, the understandability issues associated with pattern P1 have been a persistent and unresolved problem since 2022. Over these past years, at least 3 issues have been reported regarding P1. Despite the developers' efforts to resolve these issues, new corner cases continually arise, indicating that previous fixes were incomplete and unable to address all potential scenarios.

% The issues reporting the remaining 2 issues
The reports for the remaining 2 issues, namely issue \#6 and \#8, have not received any responses. This lack of engagement can be attributed to \project{CFR}'s developers generally being disinclined to solve issues reported to \project{CFR}. Nevertheless, the developers did respond to issue \#9, where they highly commended our suggestions for converting numerical literals into well-named constants. However, \project{CFR}'s developers also express their pessimistic attitude towards such strategies designs, citing concerns that they might lead to false positives, which could be annoying and downright confusing. We conjecture this may be the main barrier preventing decompiler developers from designing heuristic strategies for enhancing code understandability, even though improving the understandability of decompiled code is a common desire of all Java decompiler developers.

Therefore, in addition to understanding the current state of Java decompiler understandability in the wild, this paper also aims to explore methods that assist developers in creating and refining such strategies.  To this end, a practical way is to help developers identify cases where decompiled code exhibits significantly low understandability due to the decompilation process. % This objective also motivates the following experiments addressing our RQ3 and RQ4.

\section{Improved Metric}\label{sec:new_metric}

In this section, we introduce and validate our enhanced metric, \emph{Cognitive Complexity for Decompilation}, or Cognitive Complexity\textsuperscript{D} for short.

\subsection{Design}
The motivation behind the development of this new metric arises from the observation that neither of the existing metrics effectively assess decompiled code understandability, and the prevalence and severity of code patterns in decompiled files. Therefore, we propose Cognitive Complexity\textsuperscript{D} designed to more comprehensively assess code understandability in the context of decompilation.

Our choice of using Cognitive Complexity as the foundation for this metric is driven by several factors. Cognitive Complexity has demonstrated better performance when compared to perplexity in the assessment of decompiled code understandability. Moreover, Cognitive Complexity is able to reflect some aspects of code understandability~\cite{munoz2020empirical}. Furthermore, our study has revealed that Cognitive Complexity is effective in capturing the complexity of control flow to some extent. 
% although it exhibits limitations in assessing other aspects of code.

% By introducing and validating our Cognitive Complexity\textsuperscript{D} metric, we seek to provide a more accurate and comprehensive means of evaluating the understandability of decompiled code, addressing the limitations of existing metrics and aligning with established best practices in decompiled code understandability.

% To comprehensively assess the understandability of decompiled code, it became evident that a new metric was required. We propose Cognitive Complexity\textsuperscript{D}, which is designed to provide a more holistic evaluation, encompassing not only code structure but also expressions, formatting, and constants, thereby addressing the multi-faceted nature of decompiled code understandability.

\begin{table}[htbp]\footnotesize
\caption{Increment rules of Cognitive Complexity\textsuperscript{D}.}
\centering
    \begin{tabular}{|c|l|}
        \hline
        \textsc{\textbf{Rule}} & \textsc{\textbf{Description}} \\
        \hline
         R1 & Deeply nested structures. \\ \hline
         R2 & Omitted parenthesis in expressions with mixed operators. \\ \hline
         R3 & Excessively long lines. \\ \hline
         R4 & Omitted braces after conditional or loop statements. \\ \hline
         R5 & Inlined assignments in expressions. \\ \hline
         R6 & Numerical literals in expressions. \\
         \hline
    \end{tabular}
  %   \begin{tabular}{|c|c|l|}
		% \hline
		% \textsc{\textbf{Aspect}} & \textsc{\textbf{Rule}} & \textsc{\textbf{Description}} \\
		% \hline
  %       % Code structure  & R1 & use rules the same as Cognitive Complexity \\
  %       \multirow{2}{*}{Expression} & R1 & Mixed operators without parenthesis. \\ 
  %       \cline{2-3}
  %       	& R2 & Assignments inlined in expressions. \\
  %       \hline
  %       \multirow{2}{*}{Formatting}	& R3 & Conditional or loop statements without braces. \\
  %       \cline{2-3}
  %        & R4 & Excessively long lines. \\
  %       \hline
  %       Constant & R5 & Numerical literals in expressions. \\
  %       \hline
  %   \end{tabular}  
    \label{tab:rules-dcc}
\end{table}

% from the perspective of the control flow
% Given the effectiveness of Cognitive Complexity in identifying P1, the
Cognitive Complexity relies on a set of rules to evaluate code complexity. In our study, we have extended this approach by introducing additional rules to assess code in decompilation scenario. Six increment rules have been formulated to specifically target the six patterns, respectively. These rules are presented in \autoref{tab:rules-dcc} and are applied in conjunction with the existing Cognitive Complexity rules. %The majority of these additional rules are implemented by traversing the syntax tree of the code. % By incorporating these rules into the analysis, we are able to provide a more comprehensive evaluation of code understandability, covering a broader range of factors that influence how easily developers can comprehend decompiled code.

% in order to align with the ability of Cognitive Complexity rules to recognize patterns

%  configurable?  parameter experiments  (0.88)
Specifically, we conducted a grid search on the parameters to maximize the macro F1-score, and finally determined the additional rules as follows:

R1 increases the score by 3 when calculating the nesting increment in the context of deeply nested structures, whose nesting levels are no less than 3.

R2 adds 3 to the score when encountering each mixed operator without parentheses, excluding sequences of common arithmetic operators.

R3 adjusts the score by incorporating the length of statements with excessive number of characters. We experimented with various approaches, including increasing fixed values and increasing the quotient of the length against a predefined threshold when lines exceed this threshold. We found that the latter one performs better, which aligns with human perception: longer lines of code may pose greater challenges for developers to comprehend. We explored thresholds of 80, 100, and 120 characters respectively, ultimately determining that 120 yielded the optimal performance. Consequently, R3 integrates the quotient of the length divided by 120 into the score.

R4 increases the score by 4 when encountering omitted braces after $if$, $else$, $for$, $do$ or $while$ statements.

R5 adds 4 to the score when identifying assignments inlined within other expressions. %, such as binary expressions or method call sites.

R6 contributes 1 to the score when it identifies numerical literals in expressions, except for common numbers including -1, 0, and 1.

After the incorporation of these additional rules, we proceed to calculate Cognitive Complexity\textsuperscript{D} scores for selected files. The scores for the original files range from 1 to 201, with an average value of 14.2. With decompiler \project{CFR}, \project{Fernflower}, and \project{Jadx}, the scores of respective decompiled files range from 1 to 420, 0 to 422, and 1 to 213, yielding an average of 18.6, 21, and 15.1.
It is noteworthy that, the Cognitive Complexity\textsuperscript{D} scores of files generated by decompilers exhibit higher values than the original, which aligns with human perception, signifying a decrease in code understandability.
% during the decompilation process. 

% However, there are exceptions to this trend, specifically in the cases of \project{Jsoup} and \project{Scribe Java}. In these instances, the original files tend to exhibit higher scores, which can be attributed to a higher prevalence of P4 in the original files of these projects.

% We present the Cognitive Complexity\textsuperscript{D} scores categorized by projects and decompilers in \autoref{tab:dataset_CC_improved}. 

% \input{tables/confusion_matrix_CC_improved}
\begin{table}[tbp] \footnotesize
    \caption{Confusion Matrix and metrics of the relative understandability indicated by Cognitive Complexity\textsuperscript{D}.} 
    \label{tab:conf_matrix_CC_improved_all}
    \centering
    \begin{subtable}[htbp]{0.47\linewidth}
        \centering
        \caption{Confusion Matrix}
        \setlength{\tabcolsep}{0.35em}
            \begin{tabular}{|c|c|c|c|}
            \hline
                \diagbox{\textbf{Predict}}{\textbf{Actual}}  & Less & Equi & More \\\hline
                Less & 231 & 46 & 0 \\
                \hline
                Equi & 39 & 868 & 7 \\
                \hline
                More & 0 & 20 & 76  \\
                \hline
            \end{tabular}
            % \begin{tabular}{|c|c|c|c|}
            % \hline
            %     \diagbox{\textbf{Predict}}{\textbf{Actual}}  & Less & Equi & More \\\hline
            %     Less & 224 & 45 & 0 \\
            %     \hline
            %     Equi & 46 & 865 & 9 \\
            %     \hline
            %     More & 0 & 25 & 73  \\
            %     \hline
            % \end{tabular}
        \label{tab:conf_matrix_CC_improved}
    \end{subtable}
    \hfill
    \vspace{0.2cm}
    \begin{subtable}[htbp]{0.49\linewidth}  \footnotesize
    \centering
    \caption{Metrics}
    \setlength{\tabcolsep}{0.35em}
        \begin{tabular}{|c|c|c|c|}
        \hline
        \diagbox{\hspace{10pt}}{\hspace{10pt}}& \textbf{Precision} & \textbf{Recall}  & \textbf{F1}\\\hline
        Less & 0.83 & 0.86 & 0.84 \\\hline
        Equi & 0.95 & 0.93 & 0.94 \\\hline
        More & 0.79 & 0.92 & 0.85 \\\hline
        % Less & 0.83 & 0.83 & 0.83 \\\hline 
        % Equi & 0.94 & 0.93 & 0.93 \\\hline
        % More & 0.74 & 0.89 & 0.81 \\\hline
        \end{tabular}
    \label{tab:conf_matrix_CC_metric_improved}
    \end{subtable}
\end{table}

% \input{tables/confusion_matrix_CC_improved_per_decompiler}
% \begin{table}[htbp] \footnotesize
%     \caption{Confusion Matrix of the relative understandability of \project{CFR}/\project{Fernflower}/\project{Jadx} indicated by Cognitive Complexity\textsuperscript{D}.} \label{tab:conf_matrix_CC_impr_per_decompiler}
%     \centering
%         \setlength{\tabcolsep}{0.6em}
%             \begin{tabular}{c|c|c|c|c|}
%                 \multicolumn{2}{c}{} & \multicolumn{3}{c}{Actual} \\
%                 \cline{2-5}
%                  &  & Less & Equi & More \\
%                 \cline{2-5}
%                 \multirow{3}{*}{Pred} & Less & 94/86/47 & 17/33/20 & 0/0/0 \\
%                 \cline{2-5}
%                 & Equi & 16/13/5 & 259/270/325 & 4/5/2 \\
%                 \cline{2-5}
%                 & More & 0/0/0 & 9/4/7 & 30/18/23 \\
%                 \cline{2-5}
%             \end{tabular}
% \vspace{-0.3cm}
% \end{table}

\subsection{Evaluation}
\subsubsection{On the original dataset} 
We compute the confusion matrix and related metrics indicated by Cognitive Complexity\textsuperscript{D} utilizing the annotated dataset. We have adopted the same approach as Cognitive Complexity to transform the metrics into three categories of relative understandability. The results are presented in \autoref{tab:conf_matrix_CC_improved_all}. 
Compared to Cognitive Complexity, Cognitive Complexity\textsuperscript{D} reflects the relative understandability between original source code and code decompiled by three decompilers more accurately. The F1-scores for \textbf{\textit{Less}}, \textbf{\textit{Equi}} and \textbf{\textit{More}} are 0.84, 0.94 and 0.85 respectively, representing improvements of 0.42, 0.08, and 0.73
% CC: 0.42, 0.86, 0.12
% accuracy (79+901+7) / 1287 
% macro F1 0.47
% the overall accuracy is 91.2\%, which is a 14.6\% improvement in comparison to the accuracy exhibited by Cognitive Complexity
compared to the counterparts of Cognitive Complexity, yielding a macro F1-score of 0.88, affirming the capability of Cognitive Complexity\textsuperscript{D} in effectively identifying and categorizing all classes, thereby providing a comprehensive framework for assessing the understandability of decompiled code. 
For developers involved in the design and enhancement of decompilers, this metric offers the utility of automatically identifying deficiencies in the code generation strategies of decompilers.

\begin{table}[htbp]\footnotesize
\caption{An overview of the test set (\textsc{\textbf{ORI}}: Original. \textsc{\textbf{REC}}: Recompilable.  \textsc{\textbf{PAT}}: Able to pass tests. \textsc{\textbf{EVA}}: To be evaluated.).}
\centering
    \setlength{\tabcolsep}{0.35em}
    \begin{tabular}{|l|c|c|c|c|c|c|}
\hline
\textsc{\textbf{Project}} & \textsc{\textbf{\#ORI}} & \textsc{\textbf{\#REC}} & \textsc{\textbf{\#PAT}} & \textsc{\textbf{\#EVA}} & \textsc{\textbf{\#ORI\textsubscript{LOC}}} & \textsc{\textbf{\#EVA\textsubscript{LOC}}} \\\hline
% \midrule 
        HikariCP                 & $38$ & $19$ & $19$ & $8$  & $4837$  & $843$   \\ \hline
        Mybatis-3                & $347$& $172$& $172$& $53$ & $33843$ & $5670$  \\ \hline
        Jackson-core             & $86$ & $47$ & $47$ & $29$ & $18918$ & $7640$  \\ 
\hline
% \midrule
        \textsc{\textbf{Total}} & \textbf{$471$} & \textbf{$238$} & \textbf{$238$} & \textbf{$90$}  & \textbf{$57598$}  & \textbf{$14153$} \\
\hline
% \bottomrule
	\end{tabular}
    % \begin{tablenotes}
    %   \item[1] \textsc{\textbf{ORI}}: Original. \textsc{\textbf{REC}}: Recompilable.  \textsc{\textbf{PAT}}: Able to pass tests. \textsc{\textbf{EVA}}: To be evaluated.
    % \end{tablenotes}
	\label{tab:test_projects_overview}
\end{table}

\begin{table}[htbp] \footnotesize
    \caption{Confusion Matrix and metrics of the relative understandability indicated by Cognitive Complexity(CC), perplexity(P) and Cognitive Complexity\textsuperscript{D}(CC\textsuperscript{D}) on the test set.} 
    \label{tab:conf_matrix_test_all}
    \centering
    \begin{subtable}[htbp]{0.47\linewidth}
        \centering
        \caption{Confusion Matrix(CC)}
        \setlength{\tabcolsep}{0.35em}
            \begin{tabular}{|c|c|c|c|}
            \hline
                \diagbox{\textbf{Predict}}{\textbf{Actual}}  & Less & Equi & More \\\hline
                Less & 20 & 3 & 4 \\
                \hline
                Equi & 26 & 181 & 8 \\
                \hline
                More & 9 & 8 & 11 \\
                \hline
            \end{tabular}
        \label{tab:conf_matrix_CC_test}
    \end{subtable}
    \hfill
    \vspace{0.2cm}
    \begin{subtable}[htbp]{0.49\linewidth}  \footnotesize
    \centering
    \caption{Metrics(CC)}
    \setlength{\tabcolsep}{0.35em}
        \begin{tabular}{|c|c|c|c|}
        \hline
        \diagbox{\hspace{10pt}}{\hspace{10pt}}& \textbf{Precision} & \textbf{Recall}  & \textbf{F1}\\\hline
        Less & 0.74 & 0.36 & 0.49 \\\hline 
        Equi & 0.84 & 0.94 & 0.89 \\\hline
        More & 0.39 & 0.48 & 0.43 \\\hline
        \end{tabular}
    \label{tab:conf_matrix_CC_test_metric}
    \end{subtable}

    \begin{subtable}[htbp]{0.47\linewidth}
        \centering
        \caption{Confusion Matrix(P)}
        \setlength{\tabcolsep}{0.35em}
            \begin{tabular}{|c|c|c|c|}
            \hline
                \diagbox{\textbf{Predict}}{\textbf{Actual}}  & Less & Equi & More \\\hline
                Less & 26 & 82 & 6 \\
                \hline
                Equi & 21 & 98 & 16 \\
                \hline
                More & 8 & 12 & 1 \\
                \hline
            \end{tabular}
        \label{tab:conf_matrix_ngram_test}
    \end{subtable}
    \hfill
    \vspace{0.2cm}
    \begin{subtable}[htbp]{0.49\linewidth}  \footnotesize
    \centering
    \caption{Metrics(P)}
    \setlength{\tabcolsep}{0.35em}
        \begin{tabular}{|c|c|c|c|}
        \hline
        \diagbox{\hspace{10pt}}{\hspace{10pt}}& \textbf{Precision} & \textbf{Recall}  & \textbf{F1}\\\hline
        Less & 0.23 & 0.47 & 0.31 \\\hline 
        Equi & 0.73 & 0.51 & 0.60 \\\hline
        More & 0.06 & 0.04 & 0.05 \\\hline
        \end{tabular}
    \label{tab:conf_matrix_ngram_test_metric}
    \end{subtable}

    \begin{subtable}[htbp]{0.47\linewidth}
        \centering
        \caption{Confusion Matrix(CC\textsuperscript{D})}
        
        \setlength{\tabcolsep}{0.35em}
            \begin{tabular}{|c|c|c|c|}
            \hline
                \diagbox{\textbf{Predict}}{\textbf{Actual}}  & Less & Equi & More \\\hline
                Less & 44 & 8   & 0 \\
                \hline
                Equi & 11 & 177 & 2 \\
                \hline
                More & 0  & 7   & 21 \\
                \hline
            \end{tabular}
        \label{tab:conf_matrix_CCD_test}
    \end{subtable}
    \hfill
    \vspace{0.2cm}
    \begin{subtable}[htbp]{0.49\linewidth}  \footnotesize
    \centering
    \caption{Metrics(CC\textsuperscript{D})}
    \setlength{\tabcolsep}{0.35em}
        \begin{tabular}{|c|c|c|c|}
        \hline
        \diagbox{\hspace{10pt}}{\hspace{10pt}}& \textbf{Precision} & \textbf{Recall}  & \textbf{F1}\\\hline
        Less & 0.85 & 0.80 & 0.82 \\\hline 
        Equi & 0.93 & 0.92 & 0.93 \\\hline
        More & 0.75 & 0.91 & 0.82 \\\hline
        \end{tabular}
    \label{tab:conf_matrix_CCD_test_metric}
    \end{subtable}
\end{table}

\subsubsection{With updated dataset} 

To further validate the effectiveness of Cognitive Complexity\textsuperscript{D}, we extended our dataset by incorporating additional popular Java open-source projects from GitHub, as detailed in \autoref{tab:test_projects_overview}. We repeated experiments as the preceding steps. The corresponding results including confusion matrices and metrics are displayed in \autoref{tab:conf_matrix_test_all}. 
\autoref{tab:conf_matrix_CC_test_metric} shows the results indicated by Cognitive Complexity on the test set. Similar to \autoref{tab:conf_matrix_CC_metric_ori}, although it has a precision of 0.74 in predicting files demonstrating \textbf{\textit{Less}} understandability, it exhibits low values on F1-scores for the \textbf{\textit{Less}} and \textbf{\textit{More}} category. Likewise, perplexity encounters the same issues as reflected in its outcomes on the original dataset, as described in \autoref{subsec:rq4}.
% as depicted in \autoref{tab:conf_matrix_ngram_test_metric} and \autoref{tab:conf_matrix_ngram_metric_ori}.

The results presented in \autoref{tab:conf_matrix_CCD_test_metric} indicate that Cognitive Complexity\textsuperscript{D} demonstrates a superior and more balanced performance in predicting understandability categories compared to Cognitive Complexity and perplexity across evaluated metrics. Note that the F1-score for the \textbf{\textit{Less}}, \textbf{\textit{Equi}} and \textbf{\textit{More}} category indicated by Cognitive Complexity\textsuperscript{D} are 0.82, 0.93 and 0.82, which have improvement of 0.33, 0.04 and 0.39 respectively than those indicated by Cognitive Complexity. These improvements highlight the robustness of Cognitive Complexity\textsuperscript{D} in providing a more accurate and reliable assessment of code understandability across different categories.
In addition, Cognitive Complexity\textsuperscript{D} achieves a macro F1-score of 0.86, which is closely aligned with the macro F1-score of 0.88 obtained on the original dataset. This consistency underscores the effectiveness of Cognitive Complexity\textsuperscript{D} in accurately predicting changes in code understandability generated by decompilers.

\begin{figure}[htbp]
\centering
    \begin{subfigure}{.22\textwidth}
        \begin{lstlisting}[style=Java, basicstyle=\scriptsize]
if (condition1) {
    ...
    
    while (condition2) {
        ...
        
    }
    ...
    
}
// end
        \end{lstlisting}
    \vspace{0.1cm}
        \caption{Original}\label{lst:otherNested-ori}
    \end{subfigure}
    \hspace{0.1cm}
    \begin{subfigure}{.22\textwidth}
        \begin{lstlisting}[style=Java, basicstyle=\scriptsize]
if (condition1) {
    ...
    @@while (true) {@@
        if (condition2) {
            ...
        } else {
            ...
            return;
        }
    @@}@@
}
        \end{lstlisting}
    \vspace{0.1cm}
        \caption{\project{Jadx}}\label{lst:otherNested-jadx}
    \end{subfigure}
\caption{Excerpts of \texttt{com.\allowbreak{}fasterxml.\allowbreak{}jackson.\allowbreak{}core.\allowbreak{}util.\allowbreak{}DefaultIndenter} original and decompiled by \project{Jadx}.}
\label{fig:otherNested}
\end{figure}

Upon further analysis of the test set, we found that all the six patterns can be observed in the decompiled code. 
In addition, there exist alternative manifestations of P1. A typical example is shown in \autoref{fig:otherNested} from the project \project{Jackson-core}. Code generated by \project{Jadx} uses $while(true)$ instead of $while$ with conditions as the original source code, which increases the level of nesting, leading to deeply nested code structures i.e., pattern P1. Furthermore, similar code is also observed in code generated by Jadx of the version after the improved commit \#3 for P1\footnote{\url{https://github.com/skylot/jadx/commit/2d5c0}\label{improved_commit}}, and we have reported this issue to \project{Jadx}\footnote{\url{https://github.com/skylot/jadx/issues/2180}}. This indicates the challenges in comprehensively resolving understandability issues, highlighting the error-prone nature of the process. It also demonstrates that Cognitive Complexity\textsuperscript{D} can consistently and effectively help identify decompiled code with significantly decreased understandability compared to its original counterpart. Therefore, our Cognitive Complexity\textsuperscript{D}, we believe, can be utilized to further improve Java decompilers.

\subsubsection{With updated Java decompilers}

\begin{figure}[htbp]
    \centering
    \begin{minipage}[]{0.48\linewidth}
        \centering
        \includegraphics[width=\linewidth]{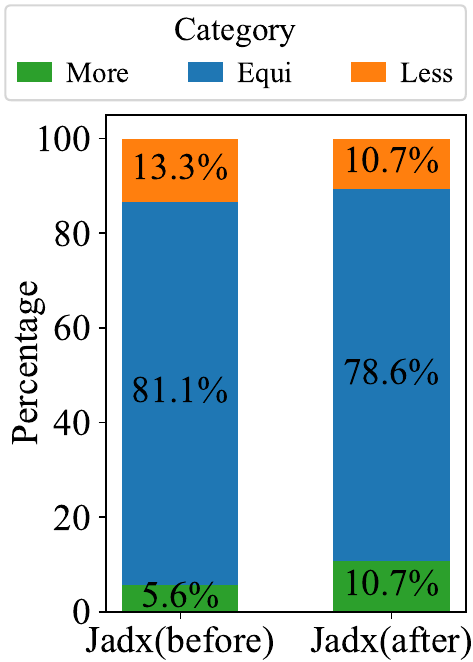}
        % \caption{(a) Relative understandability of the decompiled files}
        \subcaption{Relative understandability.}
        % \caption{An overview of relative understandability of the decompiled files when  compared to their corresponding original ones}
    \end{minipage}
    \begin{minipage}[]{0.48\linewidth}
        % \centering
        % \begin{table}[tbp] \footnotesize
        %     \caption{Metrics of the relative understandability indicated by Cognitive Complexity\textsuperscript{D} for \project{Jadx} before and after the commit.}
        %     \label{tab:}
            % \centering
            \begin{subtable}[htbp]{\linewidth}  \footnotesize
            % \centering
            % \captionsetup{labelformat=empty} % 这里去掉标签前缀
            % \caption{(b) Metrics indicated by Cognitive Complexity\textsuperscript{D} for \project{Jadx} before the commit.}
            \subcaption{Metrics indicated by Cognitive Complexity\textsuperscript{D} (before).}
            % \caption{Metrics of the relative understandability indicated by Cognitive Complexity\textsuperscript{D} for \project{Jadx} before and after the commit.}
            % \caption{Before}
            \setlength{\tabcolsep}{0.35em}
                \begin{tabular}{|c|c|c|c|}
                \hline
                \diagbox{\hspace{10pt}}{\hspace{10pt}}& \textbf{Precision} & \textbf{Recall}  & \textbf{F1}\\\hline
                Less & 0.79 & 0.80 & 0.79 \\\hline
                Equi & 0.96 & 0.94 & 0.95 \\\hline
                More & 0.72 & 0.95 & 0.82 \\\hline
                \end{tabular}
            \label{tab:conf_matrix_CC_metric_jadx_before}
            % \hspace{0.5cm}    
            \medskip
            \end{subtable}
            \begin{subtable}[htbp]{\linewidth}  \footnotesize
            \centering
            \subcaption{Metrics indicated by Cognitive Complexity\textsuperscript{D} (after).}
            \setlength{\tabcolsep}{0.35em}
                \begin{tabular}{|c|c|c|c|}
                \hline
                \diagbox{\hspace{10pt}}{\hspace{10pt}}& \textbf{Precision} & \textbf{Recall}  & \textbf{F1}\\\hline
                Less & 0.80 & 0.80 & 0.80 \\\hline
                Equi & 0.97 & 0.94 & 0.96 \\\hline
                More & 0.84 & 0.98 & 0.91 \\\hline
                \end{tabular}
            \label{tab:conf_matrix_CC_metric_jadx_after}
            \end{subtable}
        % \end{table}
    \end{minipage}
    \caption{An overview of comparison of code decompiled by \project{Jadx} before and after the improved commit.}
    \label{fig:overall_jadx}
\end{figure}

It is natural that existing Java decompilation techniques are continuously  evolving, with various heuristics strategies gradually taken. Investigating  whether our Cognitive Complexity\textsuperscript{D} is able to detect the understandability changes in the decompilers' version evolution is also of great interest. To explore this, we reproduced the experiments for Research Question 3 (RQ3) using two versions of the \project{Jadx} decompiler: one just before and one just after the fix for the understandability issue \#3 for P1. % \textsuperscript{\ref {improved_commit}}%todo\footnote{\url{https://github.com/skylot/jadx/commit/2d5c0}}. 

The results, depicted in~\autoref{fig:overall_jadx}, show the decompilation outputs of the old and new versions of \project{Jadx}, along with the metrics for relative understandability assessed by our Cognitive Complexity\textsuperscript{D} on two datasets. Upon careful examination, we observed code changes in 30 out of 429 files, resulting in 29 cases where code understandability significantly improved (from \textbf{\textit{Less}} to \textbf{\textit{Equi}}/\textbf{\textit{More}}, or from \textbf{\textit{Equi}} to \textbf{\textit{More}}). Given the only difference between the two versions of \project{Jadx} is the fix for issue \#3, we conclude that this fix did generally improve the decompilation understandability of \project{Jadx}. Nevertheless, there still remain one case where a decrease in understandability was observed.

% a significant decrease in understandability (from \textbf{\textit{Equi}} to \textbf{\textit{Less}}) was observed.

% Among the 429 files, changes in Cognitive Complexity\textsuperscript{D} are observed in 51 files. Among them, 42 files exhibited a decrease after the commit, while 9 files experienced an increase.

\begin{figure}[htbp]
\centering
    \begin{subfigure}{.22\textwidth}
        \begin{lstlisting}[style=Java, basicstyle=\scriptsize]
if (condition1) {
    ...
    
} else if (condition2) {
    ...
    
} else if (condition3) {
    ...
    
} else if (condition4) {
    ...
    
}
        \end{lstlisting}
%         \begin{lstlisting}[style=Java, basicstyle=\scriptsize]
% if (condition1) {
%     ...
% } else if (condition2) {
%     ... return false;
% } else {
%     [...]
% }

% ... return true;
%         \end{lstlisting}
    \vspace{0.1cm}
        \caption{Before}\label{lst:playSoundCommand-before}
    \end{subfigure}
    \hspace{0.1cm}
    \begin{subfigure}{.22\textwidth}
        \begin{lstlisting}[style=Java, basicstyle=\scriptsize]
if (condition1) {
    ...
@@} else {@@
    if (condition2) {
        ...
    }
    if (condition3) {
        ...
    }
    if (condition4) {
        ...
    }
@@}@@
        \end{lstlisting}
%         \begin{lstlisting}[style=Java, basicstyle=\scriptsize]
% if (condition1) {
%     ...
% @@} else {@@
%     if (condition2) {
%         ... return false;
%     }
%     [...]
% @@}@@
% ... return true;
%         \end{lstlisting}
    \vspace{0.1cm}
        \caption{After}\label{lst:playSoundCommand-after}
    \end{subfigure}
\caption{Excerpts of \texttt{org.\allowbreak{}apache.\allowbreak{}commons.\allowbreak{}imaging.\allowbreak{}formats.\allowbreak{}tiff.\allowbreak{}datareaders.\allowbreak{}BitInputStream} decompiled by \project{Jadx} before and after the improved commit.}
\label{fig:playSoundCommand}
\end{figure}

% org.\allowbreak{}bukkit.\allowbreak{}command.\allowbreak{}defaults.\allowbreak{}PlaySoundCommand}

The example is presented in \autoref{fig:playSoundCommand}, where the \code{condition2}, \code{condition3} and \code{condition4} $if$ statements are nested within the $else$ block of the \code{condition1} $if$ statement in \project{Jadx}'s new version, rather than being treated as $else\ if$ blocks as in the old version. The decompiled code in the new version features more deeply nested conditional statements, % i.e., pattern P1,  potential
and therefore, are more difficult to understand. This fix, which intended to address the understandability issue of P1, however, introduced more code snippets of P1 in decompilation in turn. This real-world example underscores the challenges inherent in developing heuristic strategies to enhance decompilation understandability. It highlights that the design of such strategies is a long-term and error-prone process, necessitating a reliable metric to assist developers in continuously revealing understandability decrements due to decompilation.
Our Cognitive Complexity\textsuperscript{D}, as shown in~\autoref{tab:conf_matrix_CC_metric_jadx_before} and~\autoref{tab:conf_matrix_CC_metric_jadx_after}, has achieved comparable precision between the dataset of the two versions of \project{Jadx}. % Moreover, Cognitive Complexity\textsuperscript{D} has shown comparable precision across the datasets from the two versions of \project{Jadx}. 
Furthermore, Cognitive Complexity\textsuperscript{D} successfully recognized the case of abnormal understandability decrement, thus proving its utility in aiding developers to refine and improve their decompilers in terms of code understandability.

\section{Threats to Validity}\label{sec:threat}
In this section, we discuss possible threats to the validity of our findings and show how we mitigate them.
\subsection{Internal Validity} These threats concern the extent to which the observed effects in a study can be attributed to the independent variable. A potential internal threat stems from our decision to engage three students in the annotation process for assessing the relative understandability of decompiled code. Code understandability is inherently influenced by various human factors. To address this inherent subjectivity, we mitigate these influences by instructing the annotators to familiarize themselves with Java coding conventions prior to the annotation task, and asked them to base their assessments on the conventions.
% review: how to evaluate their quality/ability to annotate

\subsection{External Validity} These threats relate to the generalizability of research findings to other contexts. The potential external threat arises from the selection of three decompilers, namely \project{CFR}, \project{Jadx}, and \project{Fernflower}. These decompilers were chosen due to their popularity as open-source tools, as evidenced by our user survey. Our study has revealed that most of the code patterns that change the understandability are consistently observed across multiple decompilers. This observation underscores the universality of these code patterns generated by Java decompilers.

% \subsection{Construct Validity}

\section{Related Work}\label{sec:related_work}
In this section, we present other research work related to our work, principally concentrating on empirical studies of Java decompilation and code understandability.
\subsection{Studies of Java Decompilation}

Java decompilers serve as tools for the transformation of bytecode back to human-readable Java code. Harrand et al.~\cite{harrand2020java} conducted an extensive investigation on eight Java decompilers from the perspective of syntactic correctness, syntactic similarity and semantic equivalence. Mauthe et al.~\cite{mauthe2021large} performed a large-scale study focused on Android App decompilation. % Their research primarily concentrated on the evaluation of decompilation success rates. 
Naeem et al.~\cite{naeem2007metrics} presented metrics to evaluate the effectiveness of decompilers and obfuscators, including program size, conditional statements, abrupt control flow directives and conditional complexity. We conduct an analysis focused on understandability of code generated by Java decompilers.
% assess Java decompilers from the perspective of decompiled code understandability.
% The latter three are taken into account by rules of Cognitive Complexity in Cognitive Complexity\textsuperscript{D}. We find that there is a positive monotonic relationship between Cognitive Complexity\textsuperscript{D} and token size.

\subsection{Studies of Code Understandability} % and Code Readablity

Code understandability is challenging to quantify. A prior study~\cite{scalabrino2019automatically} has revealed that there is limited correlation between individual traditional metrics and source code understandability, while combining metrics improves the correlation. Code readability has a profound impact on decompiled code understandability. Buse et al.~\cite{buse2009learning} proposed a method to model code readability with machine-learning algorithms based on a simple set of code features. In a subsequent development, Posnett et al.~\cite{posnett2011simpler} constructed a simpler model based on size metrics and Halstead metrics~\cite{halstead1977elements}. Scalabrino et al.~\cite{scalabrino2018comprehensive,scalabrino2016improving} combined structural aspects and textual features to devise a model for code readability. Fakhoury et al.~\cite{fakhoury2019improving} conduct a study on measuring source code readability improvements, concluding that existing readability models inadequately capture such improvements. % , while some additional metrics are able to. 
In our study, we aim to evaluate decompiled code understandability using a single metric. We conduct a manual analysis of decompiled files and introduce Cognitive Complexity\textsuperscript{D} as a metric to assess the code understandability in the context of decompilation.

Johnson et al.~\cite{johnson2019empirical} have demonstrated that minimizing nesting decreases the time developers expend reading and understanding source code. Casalnuovo et al.~\cite{casalnuovo2020programmers} have explored the relationship between surprisal of language models and programmer preference on semantic equivalent expressions. %, and find that programmers prefer more predictable variants.
An examination~\cite{oliveira2023systematic} into the impact of formatting elements demonstrated that format styles, indentations, block delimiters and long or complex code lines have impact on code legibility. A survey~\cite{dos2018impacts} that assesses some Java coding practices also indicates that limiting line lengths has a positive impact on readability. Our metric takes into account recurring patterns extracted from decompiled files, which have been validated as influential factors affecting code understandability.

% Program Comprehension and Code Complexity Metrics: An fMRI Study?

% In our study, code patterns that decrease the code understandability can also be observed in source code. The relationship between original source code and decompiled code in decompilation is similar to the relationship between code after and before refactorings, because they are both semantic equivalent code. Although our metric is proposed to assess decompiled code understandability, it can also be used to evaluate source code readability improvements during refactoring and maintenance from the perspective of code structure, expression, formatting and constant.

% casalnuovo2020programmers  In other contexts, we hope that as these effects are better understood, it may be possible to use surprisal as a method to guide automated tools modifying code to be more easily understandable by humans without altering its computational meaning. Such tools could include style recommendations or parts of the code editor. Finally, methods of measuring readability and comprehension of code may support more effective teaching of coding, establishing good code writing practices that are more understandable to existing programmers.

\section{Conclusion}\label{sec:conclusion}

In this work, we have conducted the first empirical study on the domain of code understandability of Java decompilation. Our user survey found that Java developers and researchers consider the understandability of Java decompilation to be as important as its correctness. Furthermore, understandability issues of decompiled code are more frequently encountered than actual decompilation failures. Our research entailed a comprehensive manual analysis of 429 sets of Java files. To ascertain metrics for assessing decompiled code understandability, we employed Cognitive Complexity and the perplexity of n-gram language models. Our findings indicate that neither of these metrics is entirely suitable for this purpose. We have identified six prevailing code patterns that do harm to the understandability of decompiled code, and discuss the prevalence and severity of them. Ultimately, we have introduced Cognitive Complexity\textsuperscript{D} as a metric to evaluate decompiled code understandability. Our metric has been validated, exhibiting a macro F1-score of 0.88 on the original dataset, and 0.86 on the test set.

% Furthermore, it performs admirably in the evaluation of the decompilers under scrutiny, demonstrating high recall rates across various classes of code.

\section{Data Availability}\label{sec:dataavailable}
All the experimental results and the tools for assessing the understandability of decompiled code with Cognitive Complexity, perplexity, and Cognitive Complexity\textsuperscript{D} are publicly available at
\url{https://doi.org/10.5281/zenodo.11474285}.

% Nonetheless, our survey results suggest that decompilation approaches could go much further to improve readability, leading to a lot of potential for future research while still retaining semantic equivalence.

% \begin{appendices}
%   \section{Questionnaire} % 

% \begin{itemize}
%     \item[1.] What is your highest level of education?
%         % \begin{description}
%         %     \item[a.] High School
%         %     \item[b.] Bachelor's Degree
%         %     \item[c.] Master's Degree
%         %     \item[d.] PhD
%         % \end{description}
%         \begin{itemize}
%             \item[\single] High School
%             \item[\single] Bachelor's Degree
%             \item[\single] Master's Degree
%             \item[\single] PhD
%         \end{itemize}

%     \item[2.] How much experience do you have in Java programming?
% 1\ \ \single\ \ \ \single\ \ \ \single\ \ \ \single\ \ \ \single\ \ \ \single\ \ \ \single\ \ \ \single\ \ \ \single\ \ \ \single\ \ 10

%     \item[3.] How often do you engage in programming activities?

% \end{itemize}

% \end{appendices}

\bibliographystyle{ieeetr}
\balance
\bibliography{main}

\end{document}